# Heun-Polynomial Representation of Regular-at-Infinity Solutions for the Basic SUSY Ladder of Hyperbolic Pöschl-Teller Potentials Starting from the Reflectionless Symmetric Potential Well

G. Natanson
ai-solutions Inc.
2232 Blue Valley Dr.
Silver Spring MD 20904
U.S.A.
greg_natanson@yahoo.com

It is shown that the regular-at-infinity (R@∞) solution of the 1D Schrödinger equation with the hyperbolic Pöschl-Teller (h-PT) potential $s(s-1)sh^{-2}r-(n-s+2)(n-s+1)ch^{-2}r$, where s and n are positive integers, is expressible in terms of a n-order Heun polynomial $Hp_n[y;\kappa;s]$ in $y\equiv thr$ at an arbitrary negative energy $-\kappa^2$. It was proven that the Heun polynomials in question form a subset of generally complex Lambe-Ward polynomials corresponding to zero value of the accessory parameter. Since the mentioned solution expressed in the new variable y has an almost-everywhere holomorphic (AEH) form it can be used as the factorization function (FF) for 'canonical Liouville-Darboux transformations' (CLDTs) to construct a continuous family of 'shape-invariant' rational potentials $^1V[y;s,n|\kappa_1]$ exactly-solvable by the so-called 'Heun-seed' (HpS) Heine polynomials. There are also two ($\dagger_+ = \mathfrak{a}$ or $\mathfrak{a}'$) sequences of infinitely many rational potentials $^1V[y;s,n|\dagger_+,m]$ generated using CLDTs with nodeless regular-at-origin (R@O) AEH FFs.

Keywords: Exactly solvable potential, quantized by polynomials potential, Heun polynomials

## 1. Introduction

It was Brattacharjie and Sudarshan [1] (see also [2]), who first pointed to a tremendous power of Darboux transformations [3, 4] for constructing new families of solvable potentials. Their work also brought author's attention [5-8] to an obscure exercise in Ince' textbook [9], which suggested the reader to use a sequence of the Darboux transformation (DT) for constructing the *general* solution of the 1D Schrödinger equation with the trigonometric potential

$$V_D(x;m,n) = \frac{m(m-1)}{sin^2 x} - \frac{n(n-1)}{cos^2 x} \tag{1.1}$$

for positive integers m and n. At those ancient times (with no computers) we were focused on a search for new exactly solvable potentials and it immediately became clear that similar solutions in elementary functions can be also constructed for the hyperbolic version of the Pöschl-Teller (h-PT) potential [10]

$$V_{s,t}(r) = \frac{s(s-1)}{sh^2 r} - \frac{(s+t)(s+t-1)}{ch^2 r}, \tag{1.1*}$$

where t and s are nonnegative integers. A certain advantage of the hyperbolic potential is that it remains finite for any positive values of r, whereas its trigonometric counter-part is infinitely large at both ends of a finite interval. While analyzing regular-at-infinity (R@∞) solutions constructed using the given sequence of DTs with the regular-at-origin (R@O) 'basic' FFs (in our current terms [11]) the author ran into a surprising result: the R@∞ solution at the energy $\varepsilon = -\kappa^2$ for any positive integers s and t are expressible in terms of 'exotic' polynomials $P_{s,t}(y;\kappa)$ with coefficients satisfying three-term recurrence relations.

Note that the first term in the right-hand side of (1.1*) vanishes at s = 0 and 1 and below we use this dualism to distinguish between two types of the boundary conditions: the reflectionless [12] symmetric Rosen-Morse (sym-RM) potential $V_{0,t}(x)$ quantized on the line $-\infty < x < +\infty$ (s = 0) and its radial reduction $V_{1,t}(r)$, with r > 0. Though the fact that potentials $V_{0,t}(x)$ for $t > 1$ form a reflectionless discrete subset of RM potentials [13] has no direct relation to our discussion we found the term 'reflectionless' convenient to unambiguously distinguish this discrete subset of our primary interest from all other sym-RM potentials with real values of the parameter t.

It was proven in [5] that the R@∞solution of Schrödinger equation with the reflectionless sym-RM potential is expressible in terms of an Jacobi polynomial at an arbitrary energy $\varepsilon = -\kappa^2$. (Several years ago similar elementary-function solutions were reported by Lekner [14].)

An author's renewed interest in this problem was stimulated by recent works of Matveev and Gaillard [15, 16] who re-introduced hyperbolic potential (1.1*) in context of integrable functional-difference deformations of the related difference Schrödinger equation. A thorough analysis of our old works [6-8] revealed that the Fuschian equation for the mentioned 'exotic' polynomials is converted into the Heun equation [17, 18] by a trivial gauge transformation so that the R@∞ solution of our interest is expressible in terms of Heun polynomials in $y = th\,r$.

The paper is organized as follows. In next section we discuss the necessary and sufficient conditions for existence of Heun polynomials treating the latter as a subset of generally complex Lambe-Ward polynomials [19] corresponding to real values of the accessory parameter. We point to three classes of SUSY partners of r-GRef potentials [20, 21, 11] which can be either exactly or conditionally-exactly[x)] quantized by Heun polynomials constructed in such a way. Each of these three classes will be studied in a separate publication. In this paper we solely focus on application of Lambe-Ward' equations to the Heun equation describing R@∞ solutions of the Schrödinger equation with exactly-solvable-by-Heun-polynomials (Hp-ES) h-PT potential (1.1*). Note that we distinguish between the terms 'exactly solvable' and 'exactly quantized' – the h-PT potential is exactly quantized by Jacobi polynomials for any nonnegative values of the

---

x) We use Junker and Roy's [22, 23] term 'conditionally exactly' to stress that the positions of singular points of the resultant potential depend on other parameters.

parameters t and s whereas we are interested only in a discrete subset formed by nonnegative integers.

As demonstrated in Section 3 the R@∞ solutions in question are expressible in terms of Heun polynomials form a subset of generally complex Lambe-Ward polynomials corresponding to zero value of the accessory parameter. Since the mentioned solution of the associated rational canonical Sturm-Liouville equation (RCSLE) has an almost-everywhere holomorphic (AEH) form it can be used as the factorization function (FF) for 'canonical Liouville-Darboux transformations' (CLDTs) to construct a continuous family of 'shape-invariant' rational potentials ${}^{1}V[y;s,n|\kappa_1]$ conditionally-exactly-solvable by Heine polynomials [24, 25] of a special type referred to below as 'Hp-seed' (HpS) Heine polynomials. Besides an analysis of the latter potential, Section 4 additionally exploits two sequences of infinitely many rational potentials ${}^{1}V[y;s,n|\mathfrak{t}_+,m]$, with $\mathfrak{t}_+ = \mathfrak{a}$ or $\mathfrak{a}'$, which are generated by means of CLDTs with nodeless R@O AEH FFs and therefore are also conditionally-exactly-solvable by HpS Heine polynomials.

## 2. General family of rational potentials solvable via Heun functions

Let us start our analysis from the RCSLE with four regular singular points (including infinity):

$$\left\{\frac{d^2}{dz^2} + I[z;\varepsilon\,|\,{}_1\mathsf{H}]\right\}\Phi[z\,|\,{}_1\mathsf{H}] = 0\,, \tag{2.1}$$

where the Bose invariant [26]

$$I[z;\varepsilon\,|\,{}_1\mathsf{H}] \equiv I^{o}[z;z_{out}\,|\,\bar{\lambda}_o, O_1^o] + {}_1\wp[z;T_{K_*}]\,\varepsilon \tag{2.2}$$

is represented by a polynomial fraction (PF) with three second-order poles:

$$I^{o}[z;z_{out}\,|\,\bar{\lambda}_o, O_1^o] = \sum_{r=0}^{2} \frac{1-\lambda_{o;r}^2}{4(z-e_r)^2} + \frac{O_1^o[z]}{4\Pi_3[z;\bar{e}]} \tag{2.3}$$

and

$$ {}_1\wp[z;T_{K_*}] = \frac{T_{K_*}[z]}{4\Pi_3^2[z;\bar{e}]} \quad (K_* \leq 4). \tag{2.4} $$

Here

$$ \Pi_N[z;\bar{e}] \equiv \prod_{r=0}^{N-1} (\eta - e_r) \ , \tag{2.5} $$

with $\bar{e} \equiv (0, 1, z_{out})$. The numerator of PF (2.4) is referred to as the tangent polynomial (TP), with $K_*$ standing for its order.

The change of variable $z(x)$ determined via the first-order differential equation

$$ z'(x) = \wp^{-1/2}[z(x);\bar{e};T_{K_*}], \tag{2.6} $$

converts RCSLE (2.1) into the Schrödinger equation with the Liouville potential [24]:

$$ V[z(x)\,|\,H] = -\wp^{-1}[z(x);T_{K_*}]\, I^o[z;\bar{e}\,|\,Q^H] - \tfrac{1}{2}\{z(x), x\}\,, \tag{2.7} $$

where prime is used for the derivative with respect to x and the symbol $\{\xi, x\}$ denotes the so-called Schwartz derivative (see, i.g., [27]).

Last year Batic, Williams, and Nowakowski [28] broadened our results [20] for a general family of rational potentials solvable by hypergeometric and confluent hypergeometric (*c*-hypergeometric) functions -- referred in our recent works [11, 21] as *r*- and *c*-Gauss reference (*r*- and *c*-GRef) potentials by considering rational potentials solvable via Heun and *c*-Heun functions. (As a matter of fact, they also considered rational potentials solvable via biconfluent and triconfluent Heun functions [18] but the associated SLEs are beyond the scope of this series of publications.) It is worth mentioning in this connection that some steps in extension of Bose's paper [26] to potentials solvable by Heun functions have been made by Lemieux and Bose [29] two years before publication of our paper [20]. However, similarly to Bose's work, these authors again restricted their analysis solely to TPs whose zeros coincide with one of singular points of the RCSLE in question.

It is crucial that any change of variable resulting in a potential solvable via Heun or *c*-Heun functions must satisfy first-order differential equation (2.6), with the rational density function defined via (2.4). The classical example is given by the Lame equation which is obtained from (2.1) by setting $\lambda_{o;0} = \lambda_{o;1} = \lambda_{o;2} = 1/2$ in (2.3) and using the Weierstrass elliptic function as the new variable z(x). One can directly see from (9) – (11) in [30] that

$$T_3(z) = 1/4\, z(1-z)(z-z_{out})\ . \tag{2.8}$$

in this case.

However, in this paper we are only in rational potentials which are quantized by Heun polynomials, not just solvable via Heun functions or their *c*-counterparts. To find such analytically quantized solutions it is convenient to parameterize the TP as

$$T_{K*}[z] = [c_0(z-1)^2 + c_1 z^2](z-z_{out})^2 + c_2 z^2(z-1)^2 + d_H\, z(z-1)(z-z_{out})(z-z_o), \tag{2.9}$$

where

$$c_r \equiv \frac{T_{K_*}[e_r]}{(e_r - e_{r'})^2 (e_r - e_{r''})^2} \tag{2.9′}$$

(with r, r′, and r″used to denote three distinct indexes equal to 0, 1, or 2),

$$d_H \equiv T_{K*;4} - \sum_{r=0}^{2} c_r \tag{2.9″}$$

(with $T_{K*;4}$ standing for the leading coefficient of the polynomial $T_{K*}[z]$), and

$$z_o \equiv z - \frac{T_{K*}[z] - [c_0(z-1)^2 + c_1 z^2](z-z_{out})^2 + c_2 z^2 (z-1)^2}{d\, z(z-1)(z-z_{out})}, \tag{2.9‴}$$

so that Bose invariant (2,2) takes the form:

$$I[z;\varepsilon\,|\,{}_1H] \equiv \sum_{r=0}^{2} \frac{1-\lambda^2(\varepsilon;\lambda_{o;r};c_r)}{4(z-e_r)^2} + \frac{O_1^o[z;\varepsilon\,|\,{}_1H]}{4\Pi_3[z;\bar{e}]}, \tag{2.10}$$

where we set

$$\lambda(\varepsilon;\lambda_o;c) \equiv \sqrt{\lambda_o^2 - c\,\varepsilon} \tag{2.11}$$

and

$$O_1^0[z;\varepsilon\,|\,_1H] \equiv O_1^0[z] - d_H(z - z_0)\varepsilon. \tag{2.12}$$

We then use the gauge transformation

$$\Phi[z;\varepsilon\,|\,_1H] = \Theta[z;\bar{e};\sigma_0\lambda(\varepsilon;\lambda_{o;0};c_0),\lambda(\varepsilon;\sigma_1\lambda_{o;1};c_1),\sigma_2\lambda(\varepsilon;\lambda_{o;2};c_2)] \times F[z;\varepsilon\,|\Phi[\xi;\varepsilon\,|\,_1H;\bar{\sigma}], \tag{2.13}$$

where

$$\Theta[z;\bar{e};\bar{\lambda}] \equiv \prod_{r=0}^{2} |z - e_r|^{\frac{1}{2}(\lambda_r+1)}, \tag{2.14}$$

which results in the following Heun equation written in the 'asymmetrically reduced form' [31]:

$$\left\{\frac{d^2}{dz^2} + 2\sum_{r=0}^{2} \frac{\rho_r(\varepsilon\,|\,_1\mathbf{H};\bar{\sigma})}{z - e_r}\frac{d}{dz} + \frac{C_1[z;\varepsilon\,|\,_1\mathbf{H};\bar{\sigma}]}{\Pi_3[z;\bar{e}]}\right\}{}_1\mathrm{Hf}[z;\varepsilon\,|\,_1\mathbf{H};\bar{\sigma}] = 0, \tag{2.15}$$

where the 'canonical' characteristic exponents are given by the conventional formula

$$\rho_r(\varepsilon\,|\,_1\mathbf{H};\bar{\sigma}) \equiv \tfrac{1}{2}[1 + \sigma_r\lambda(\varepsilon;\lambda_{o;r};c_r)], \tag{2.16}$$

and

$$C_1[z;\varepsilon\,|\,_1\mathbf{H};\bar{\sigma}] = \tfrac{1}{4}O_1[z;\varepsilon\,|\,_1\mathbf{H}] + \Sigma_1[z;\bar{e},{}_1\bar{\rho}(\varepsilon\,|\,_1\mathbf{H};\bar{\sigma})]\cdot \tag{2.17}$$

The second term in the right-hand side of (2.17) is the generic polynomial

$$\Sigma_{N-2}[z;\bar{e},\bar{\rho}] = \Pi_N[z;\bar{e}] \sum_{r=0}^{N-1} \frac{\rho_r}{z - e_r} \sum_{r'\neq r}^{N-1} \frac{\rho_{r'}}{z - e_{r'}} \tag{2.18}$$

appearing in the conversion formula [30] from the RCSLE to appropriate Heine equations. The coefficients of polynomial (2.17) are related to the conventional parameters [18] in the Heun equation by the standard formulas

$$C_{1;0}[z;\varepsilon\,|\,{}_1\mathbf{H};\bar{\sigma}] = -q(\varepsilon\,|\,{}_1\mathbf{H};\bar{\sigma}), \tag{2.19a}$$

$$C_{1;1}[z;\varepsilon\,|\,{}_1\mathbf{H};\bar{\sigma}] = \alpha[z;\varepsilon\,|\,{}_1\mathbf{H};\bar{\sigma}]\;\beta[z;\varepsilon\,|\,{}_1\mathbf{H};\bar{\sigma}], \tag{2.19b}$$

and

$$\alpha(\varepsilon\,|\,{}_1\mathbf{H};\bar{\sigma}) + \beta(\varepsilon\,|\,{}_1\mathbf{H};\bar{\sigma}) = 2\sum_{r=0}^{2}\rho_r(\varepsilon\,|\,{}_1\mathbf{H};\bar{\sigma}) - 1 \tag{2.20}$$

$$= \sum_{r=0}^{2}\sigma_r\lambda(\varepsilon;\lambda_{o;r};c_r) + 2. \tag{2.20′}$$

Since all the exponent differences and accessory parameter are generally energy-dependent, the HRef rational potentials cannot be usually quantized by polynomials, contrary to the r-GRef potentials *exactly* quantized by Jacobi polynomials.

As proven by Lambe and Ward [19] the Heun equation has a polynomial solution $G_{n,m}[z;\bar{\rho};z_{out}]$ of order n iff

$$\alpha[z;\varepsilon\,|\,{}_1\mathbf{H};\bar{\sigma}] = -n \tag{2.21}$$

and the appropriate accessory parameter $q_{n;m}$ (m=1,…,n) coincides with one of (generally complex) eigenvalues of the following eigenproblem

$$h^{(n)}_{0,1}G^{(n,k)}_1 + \left[h^{(n)}_{0,0} - q^{(n)}_k\right]G^{(n,k)}_0 = 0; \tag{2.22$^0$}$$

$$h^{(n)}_{j,j+1}G^{(n)}_{j+1} + (h^{(n)}_{j,j} - q^{(n)}_k)G^{(n,k)}_j + h^{(n)}_{j,j-1}G^{(n,k)}_{j-1} = 0 \tag{2.22}$$

$$\text{for } j = 1,2,...,n-1;$$

$$\left[h^{(n)}_{n,n} - q^{(n)}_k\right]G^{(n,k)}_n + h^{(n)}_{n,n-1}G^{(n,k)}_{n-1} = 0, \tag{2.22′}$$

where

$$h^{(n)}_{j,j+1} = (j+1)(2\rho_1 + j)z_{out}, \tag{2.23a}$$

$$h^{(n)}_{j,j} = -j^2(z_{out}+1) - j[z_{out}(2\rho_1 + 2\rho_2 - 1) + 2\rho_1 + 2\rho_3 - 1], \tag{2.23b}$$

$$h^{(n)}_{j,j-1} = (j-n-1)[\beta_n(\bar{\rho}) + j - 1], \tag{2.23c}$$

with

$$\beta_n(\bar{\rho}) = 2\rho_1 + 2\rho_2 + 2\rho_3 + n - 1 \, . \tag{2.24}$$

In the recent series of publications [33-35] Choun claims that there are three types of Heun polynomials:

(1) $B_{n+1} = 0$, $A_{n+1} \neq 0$;
(2) $A_{n+1} = 0$, $B_{n+1} \neq 0$;
(3) $B_{n+1} = 0$, $A_{n+1} = 0$,

where the coefficients

$$A_j = -(h^{(n)}_{j,j} - q^{(n)}_k)/h^{(n)}_{j,j+1}, \; B_j = -h^{(n)}_{j,j-1}/h^{(n)}_{j,j+1} \tag{2.25}$$

are obtained from (4a) and (4b) in [33] by setting $\lambda = 0$ , $\rho_1 = \rho_2 = \rho_3 = 1/4$, $a = 0$, $b = 1$, $c = z_{out}$, $\alpha = 2n$, and

$$\beta_n(\bar{\rho}) = 1/2(\alpha + 1) = n + 1/2 \, . \tag{2.26}$$

At odds with statements in [34], the general case is represented by the Heun polynomials of type 1, not 3. On the contrary, polynomial solutions of type 3 may appear only for some very specific combinations of values of $z_{out}$ and $\bar{\rho}$ . In particular, if $\rho_2 = \rho_3$ then diagonal elements (2.23b) take the form

$$h^{(n)}_{j,j} = -j(j + 2\rho_1 + 2\rho_2 - 1))(z_{out} + 1) \tag{2.27}$$

and therefore vanish for any values of $\rho_1$ and $\rho_2$ for $z_{out} = -1$. As a result the appropriate Heun equation always has a polynomial solution for the zero accessory parameter and any even n=2m. It is an even polynomial with coefficients satisfying two-term recurrence relations. Since all the coefficients $A_j$ vanish these polynomial solutions are formally of type 3. In next Section we present some close-form examples of polynomial solutions of this type. Finally, according to Lambe-Ward's proof, solutions of type 2 may not exist at least in case of $\lambda = 0$ associated with finite-at-origin Heun functions.

In separate publications [11, 21, 32] w e have already pointed to the fact that the 'canonical Liouville-Darboux transformations' (CLDTs) as we refer to them now [11],

$$
\begin{aligned}
{}^{1}I_{\dagger,m}[z;{}_{\iota}\bar{e}\,|\,{}_{\iota}Q^{H}] &= {}_{\iota}I^{o}[\xi;{}_{\iota}\bar{e}\,|\,{}_{\iota}Q^{H}] \\
&+ 2\;{}_{\iota}\wp^{1/2}[\xi;{}_{\iota}\bar{e};T_{K_*}]\frac{d}{d\xi}\left\{\wp^{-1/2}[\xi;{}_{\iota}\bar{e};T_{K_*}]ld\,{}_{\iota}\psi_{\dagger,m}[\xi\,|\,{}_{\iota}G_{\downarrow\dagger,m}]\right\},
\end{aligned}
\tag{2.28}
$$

results in rational potentials quantized via the so-called 'Gauss-seed' [11, 21] Heine polynomials if the solution,

$$
\psi_{\dagger,m}[z\,|\,{}_{1}G_{\downarrow\dagger,m}] = {}_{1}\wp^{1/4}[z;0,1;T_{K_*}]\phi_{\dagger,m}[z\,|\,{}_{1}G_{\downarrow\dagger,m}]
\tag{2.29}
$$

of the Schrödinger equation with the GRef potential $V[\xi\,|\,{}_{\iota}G_{\downarrow\dagger,m}]$ has the almost-everywhere holomorphic form:

$$
\phi_{\dagger,m}[z\,|\,{}_{1}G_{\downarrow\dagger,m}] = \prod_{r=0}^{1}|z-r|^{\frac{1}{2}(\lambda_{r;\dagger,m}+1)}\,P_{m}^{(\lambda_{0;\dagger,m},\lambda_{1;\dagger,m})}(2z-1),
\tag{2.30}
$$

when expressed in terms of the variable $z(x)$. (By definition the sub-beam $\mathbf{B}_{\downarrow\tau}$ of the rational beam $\mathbf{B}_{\dagger}$ is formed by rational rays described by a subset of possible values of ray identifiers which may vary depending on the chosen (fixed for the given beam) TP coefficients.) It is important that the density function in question

$$
\wp[z;0,1;T_{K}] = \frac{T_{K}[z]}{4z^{2}(1-z)^{2}} \qquad (0 \le K \le 2).
\tag{2.31}
$$

does not has zeros at the outer singular points ($c_2$=0, $z_o = z_{out}$ for HRef SUSY partners of GRef potentials). Another crucial common feature of these rational beams is that all the outer singularities in the appropriate quantized-by-polynomials differential equations are apparent [9], with the exponent differences equal either to 2 or 3 and therefore belong to the class of Fuschian differential equations defined by Conjecture 1 in [36].

As proven in [35] the lowest-energy eigenfunction of the r-GRef potential generated by means of a TP with positive discriminant is always accompanied by three basis solutions

$$\phi_{\mathbf{t},0}[z\,|\,\mathcal{G}_{\downarrow\mathbf{t},0}] = \prod_{r=0}^{1} |\,z - r\,|^{\frac{1}{2}(\lambda_{r;\mathbf{t}}+1)} \tag{2.32}$$

regular at one of the ends and irregular at both singular points ($\mathbf{t}$ = $\mathbf{a}$, $\mathbf{b}$, and $\mathbf{d}$, respectively). All three solutions are nodeless by definition and lie below the ground energy levels. Therefore each of them can be used as the FF to construct a rational SUSY partner $V[z|{}^{1}_{1}\mathcal{G}^{K\Im 0}_{\mathbf{t},0}]$ of the r-GRef potential. Together with the CGK potential [38] ($\mathbf{t}$ = $\mathbf{c}$) they form a quartet of the sibling potentials quantized by polynomials. In case of the generic r-GRef potential on the line, $V[z|\mathcal{G}^{K\Im 0}_{\downarrow\mathbf{t},0}]$ these polynomials satisfy the Heine equation with 5 regular singular points (including infinity). A common remarkable feature of this quartet of rational potentials is that the outer singular points are independent of the ray identifiers. For this reason we refer to these potentials as ‘exactly quantized’ by GS Heine polynomials despite the fact that most of the ray identifiers of the four rational beams ${}^{1}_{1}\mathcal{G}^{K\Im 0}_{\mathbf{t},0}$ are restricted by some constraints.

In case of the TP with a single root $z_{out}$ we come to a rational potential quantized via Heun polynomials. There are three such cases discussed in a separate publication, together with its confluent counterparts:

a) linear TP (LTP)

$$T_1[z;1,0] = b(z - z_{out}),\ \lambda_{o,2} = 2; \tag{2.33a}$$

b) zero root TP (ZRtTP)

$$T_2[z;1,1] = az(z - z_{out}),\ \lambda_{o,2} = 2; \tag{2.33b}$$

c) double root TP (DRtTP)

$$T_2[\xi;1,0] = a(z - z_{out})^2,\ \lambda_{o,2} = 3. \qquad (2.33c)$$

As a result the outer singular point in the Heun equation coincides with $z_{out}$ and therefore is independent of the ray identifiers.

If we now fix the energy and the appropriate values of the energy-dependent exponents and vary the accessory parameter independently then there is a sequence of $n(\varepsilon;\lambda_{o;0}, \lambda_{o;1})$ of Heun polynomials however the appropriate values of the accessory parameter do not match (2.20a). Only if we select ε equal to the energy $\varepsilon_v$ of the $v^{th}$ bound state the value of the accessory parameter for the Heun polynomial of order v satisfies (2.17a), as required.

It can be proven that CLDT using a basic solution †,0as its FF converts any other basic solution †′,0into the AEH solution

$$\phi_{\dagger\dagger'}[z\,|\,{}^*G_{\dagger,0\downarrow\dagger',0}] = \wp^{-1/2}[z]\; z^{\frac{1}{2}\lambda_{0;\dagger',0}}\,(1-z)^{\frac{1}{2}\lambda_{1;\dagger',0}}\,[\,z - z_{\dagger\dagger'}(Q^G)], \qquad (2.34)$$

where

$$z_{\dagger\dagger'}(Q^G) = \frac{\lambda_{0;\dagger',0} - \lambda_{0;\dagger,0}}{\lambda_{0;\dagger',0} + \lambda_{1;\dagger',0} - \lambda_{0;\dagger,0} - \lambda_{1;\dagger,0}}. \qquad (2.35)$$

Making use of Suzko's reciprocal formula [39, 40]

$${}^*\phi_{\dagger\dagger'}[z\,|\,{}^*G_{\dagger,0\downarrow\dagger',0}] = \wp^{-1/2}[z]/\,\phi_{\dagger\dagger'}[z\,|\,{}^*G_{\dagger,0\downarrow\dagger',0}] \qquad (2.36)$$

(see Appendix B in [21] for details) the FF for the reverse CLDT can be represented as

$${}^*\phi_{\dagger\dagger'}[z\,|\,{}^*G_{\dagger,0\downarrow\dagger',0}] = z^{-\frac{1}{2}\lambda_{0;\dagger',0}}\,(1-z)^{-\frac{1}{2}\lambda_{1;\dagger',0}}\,[z - z_{\dagger\dagger'}(Q^G)]^{-1} \qquad (2.37)$$

so that the corresponding CRSLE has only four regular singular points 0, 1, $z_{\dagger\dagger'}(Q^G)$, and ∞.

One can construct a rational extension (RE) solvable via Heun functions [29] by using 'primitive' TPs

$$T_K[z;0,\aleph] = a\, z^{\aleph}(1-z)^{K-\aleph} \tag{2.38}$$

associated with 'shape-invariant' GRef potentials. The appropriate rational beams will be referred to as $\boldsymbol{H}^{K0\aleph}$ by analogy with their GRef sub-beams $\boldsymbol{G}^{K0\aleph}$. Though each potential $V[z|\boldsymbol{H}^{K0\aleph}]$ can be represented as an explicit function of x its energy spectrum cannot be obtained in the closed form, contrary to $V[z|\boldsymbol{G}^{K0\aleph}]$ . Note that we distinguish here between the terms 'rational extension' and 'rational SUSY partner'. In general, rational extensions of shape-invariant GRef potentials are not exactly quantized, contrary to the SUSY partners constructed by means of CLDTs with nodeless AEH factorization functions (FFs).

The SUSY partners in question are conditionally exactly quantized by Heun polynomials (Hp-CEQ) if the polynomial in the right-hand side of (2.21) is linear:

$$P_1[z;\boldsymbol{G}_{\downarrow\dagger,1}] = z - z_{out}(Q^G)\,. \tag{2.39}$$

As pointed to in [21] the outer root is given by the generic expression

$$z_{out}(Q^G) = \frac{\lambda_{0;\dagger,1}+1}{\lambda_{0;\dagger,1}+\lambda_{1;\dagger,1}+2} \tag{2.40}$$

For polynomial to be nodeless within the quantization interval (0,1) the exponents ${}_1\lambda_{\dagger,1}$ and ${}_1\nu_{\dagger,1}$ must satisfy the constraint

$$(\lambda_{0;\dagger,1}+1)(\lambda_{1;\dagger,1}+1) < 0 \tag{2.41}$$

so that

$$z_{out}^{-1}(Q^G) = 1 + \frac{\lambda_{1;\dagger,1}+1}{\lambda_{0;\dagger,1}+1} < 1. \tag{2.42}$$

A very remarkable example of Hp-CEQ rational potentials is given by Quene's [41, 42] rational SUSY partner of the Darboux/Pöschl-Teller (D/PT) potential [43, 10] which is quantized via $X_1$-Jacobi polynomials [44, 45] and therefore is referred to below as the $X_1$J-Q potential, to distinguish it from Odake and Sudaki's general family of rational potentials [ 46, 47] quantized via $X_m$-Jacobi polynomials [48, 49]. As a matter of fact, Quesne [41] started her analysis based on the fact that the $X_1$-Jacobi polynomials [42, 43] satisfy the Heun equation [36]. She then represented this equation in the canonical form, implicitly made the Liouville transformation, and wrote an explicit expression for the resultant ($X_1$J-Q) Liouville potential. The unique feature of the latter potential is that all the exponent differences become energy-independent due to the special choice of the TP:

$$ {}_1T_2[z;0,1] = {}_1a\, z(1-z). \tag{2.43} $$

and as a result the accessory parameter becomes energy-independent [21].

A detailed analysis of all the mentioned Hp-EQ and Hp-CEQ potentials will be presented in a separate series of publications. The only important point for this study is that the canonical characteristic exponent associated with the outer singular point is negative in each of the cited cases so that the appropriate Heun polynomials do not belong to the class of Heine-Stieltjes polynomials [24, 50, 51]. In particular the author is unaware of any general constraints imposed on zeros of the Heun polynomials appearing in the cited cases. In next Section we present a unique case of the rational potential whose eigenfunctions can be indeed expressed via Heine-Stieltjes polynomial solutions of the Heun equation.

## 3. Heun Representation for the Schrödinger equationwith the radial *r*-GRef potential

A fundamentally different type of exactly-quantized HRef potentials can be obtained by making the change of variable [11]

$$ y = \sqrt{z}\,. \tag{3.1} $$

In the GRef SLE

$$ \left\{ \frac{d^2}{dz^2} + I[z;\varepsilon\,|\,{}_1\boldsymbol{G}^{K\Im 1}] \right\} \Phi[z\,|\,{}_1\boldsymbol{G}^{K\Im 1}] = 0\,, \tag{3.2} $$

associated with the radial *r*-GRef potential [52, 53, 11] , $V[z|{}_1\boldsymbol{G}^{K\Im 1}]$. The GRef Bose invariant appearing in (3.2) is defined as follows [11]

$$I[z;\varepsilon \mid {}_1\boldsymbol{G}^{K\Im 1}] = {}_1I^{o}_{G}[z;\lambda_o,\mu_o] + {}_1\wp_{G}^{-1/2}[z;a,b]\varepsilon, \tag{3.3}$$

where[x)]

$${}_1I^{o}_{G}[z;\lambda,\mu] = \frac{1-\lambda^2}{4z^2} + \frac{1}{4(1-z)^2} + \frac{\mu^2-\lambda^2+1}{4z(1-z)} \tag{3.4}$$

and

$${}_1\wp_{G}[z;a,b] \equiv \frac{az+b}{4z(1-z)^2}. \tag{3.5}$$

The Bose invariant

$$I[y;\varepsilon \mid {}_1\boldsymbol{H}^{K\Im 1}] = {}_1I^{o}_{H}[y;\lambda_o,\mu_o] + {}_1\wp_{H}[y;a,b]\varepsilon \tag{3.6}$$

associated with the transformed RCSLE:

$$\left\{\frac{d^2}{dy^2} + I[y;\varepsilon \mid {}_1\boldsymbol{H}^{K\Im 1}]\right\}\Phi[y \mid {}_1\boldsymbol{H}^{K\Im 1}] = 0 \tag{3.7}$$

is related to (3.3) in the standard way [27]

$$I[y;\varepsilon \mid {}_1\boldsymbol{H}^{K\Im 1}] = \left(\frac{dz}{dy}\right)^2 I[y^2;\varepsilon \mid {}_1\boldsymbol{G}^{K\Im 1}] + 1/2\{z,y\}. \tag{3.8}$$

Substituting

$$\frac{dz}{dy} = 2y \tag{3.9}$$

and

$$\{z,y\} = -3/2\, y^{-2} \tag{3.10}$$

thus gives

---

x) There is a misprint in (5.1.4) in [11] -- the numerator of the last fraction should be changed for $\mu^2-\lambda^2-\nu^2+1$.

$$ {}_1I^{o}_{H}[y;\lambda_o,\mu_o] = \frac{1/4 - \lambda_o^2}{y^2} + \frac{y^2}{(1-y^2)^2} + \frac{\mu_o^2 - \lambda_o^2 + 1}{1-y^2} \tag{3.11} $$

and

$$ {}_1\wp_{H}[y;a,b] = \frac{a\,y^2 + b}{(1-y^2)^2} \tag{3.12} $$

so that $e_2 = -1$. By comparing (3.10) and (3.11) with (2,2) and (2,3), with z changed for y, we thus find;

$$ \lambda_{o;0} = 2\,\lambda_o,\ \ \lambda_{o;1} = \lambda_{o;2} = 0; \tag{3.13} $$

$$ O^{o}_{1}[y] = 2(2\lambda_o^2 - 2\mu_o^2 - 1)y; \tag{3.14} $$

$$ T_4[y] = 4y^2\,(ay^2 + b). \tag{3.15} $$

The fact that the exponent difference $\lambda_{o;\,0}$ at the singular point y=0 in CRSLE (3.7) is twice larger than the so-called [11] 'singularity index' (SI)

`

$$ \Diamond_o \equiv s_o - 1/2 = \lambda_o \tag{3.16} $$

for the radial potential $V[y^2(r)\,|\,{}_1\boldsymbol{G}^{K\Im 1}]$, with $s_o$ standing for the largest characteristic exponents at the semi-axis origin r=0, is the direct consequence of the fact that the variable y(r) is proportional to r at small r so that

$$ \lambda_{o;0} = 2\,s_o - 1\,. \tag{3.17} $$

Alternatively, parameterizing TP (3.15) according to (2.9) one finds

$$ T_4[y] = y^2\,[c_1(y^2 + 1) + d_H(y^2 - 1)], \tag{3.18} $$

where

$$ c_1 \equiv\ c_2 = a + b \tag{3.18′} $$

and

$$ d_H = 2c_1 - 4b = 2 - 4b. \tag{3.18″} $$

[One can easily verify that the right-hand side of (2.9‴) is equal to 0 if the TP in question is defined via (3.15), with z changed for y.] Since $c_1 = c_2$ the energy-dependent exponent differences for the singular points −1 and +1 coincide'. In following [11] we choose

$$c_1 \equiv a + b = 1. \tag{3.19}$$

Note that the parameter $z_o$ in the right-hand side of (2.9) vanishes so that substituting (3.14) into (2.13) gives

$$O_1[y;\varepsilon\,|\,{}_1\mathbf{H}^{K\Im 1}] = [2(2\lambda_o^2 - 2\mu_o^2 - 1) + d_H\varepsilon]y \tag{3.20}$$

so that Bose invariant

$$I[y;\varepsilon\,|\,{}_1\mathbf{H}^{K\Im 1}] = \frac{1/4 - \lambda_o^2}{y^2} + \frac{1}{4(1-y)^2} + \frac{1}{4(1+y)^2} + \frac{\lambda_o^2 - \mu_o^2 - 1/2 + 1/4\, d_H\varepsilon}{y^2 - 1} \tag{3.21}$$

has only the second-order pole at the origin. As a result the singularity at $\varsigma = 0$ disappears if one sets the exponent difference $\lambda_o$ to $1/2$ and CRSLE (3.2) turns into the differential equation with three regular singular points (including infinity).

Out of 8 possible gauge transformations eliminating the second-order poles let us consider only those

$$\Phi_{\bar\sigma}[y;\varepsilon\,|\,{}_1\mathbf{H}^{K\Im 1}] = y^{1/2 \pm \lambda_o}\left(1-y^2\right)^{1/2\left(1+\sigma\sqrt{|\varepsilon|}\right)} \mathrm{Hf}[y;\varepsilon\,|\,{}_1\mathbf{H}^{K\Im 1};\pm,\sigma,\sigma] \tag{3.22}$$

which use the same canonical characteristic exponents for the singular points −1 and 1. Namely, we choose

$$\bar\rho(\varepsilon\,|\,{}_1\mathbf{H};\pm,\sigma,\sigma)] = 1/2\left(1 \pm 2\lambda_o, 1+\sigma\sqrt{|\,\varepsilon\,|}, 1+\sigma\sqrt{|\,\varepsilon\,|}\right). \tag{3.23}$$

By definition the Heun functions in the right-hand side of (3.22) coincide with Frobenius solutions at the origin so that

$$\mathrm{Hf}[0;\varepsilon\,|\,\pm,\sigma,\sigma]=1. \tag{3.24}$$

Since $z_{out}$= -1 the Heun equation

$$\left\{\frac{d^2}{dy^2}+\left[\frac{1\pm 2\lambda_o}{y}+\frac{2(1+\sigma\sqrt{|\,\varepsilon\,|})\varsigma^2}{y^2-1}\right]\frac{d}{dy}+\frac{C_1[\varsigma;\varepsilon\,|\,\pm,\sigma,\sigma]}{y(y^2-1)}\right\}\mathrm{Hf}[y;\varepsilon\,|\,\pm,\sigma,\sigma]=0, \tag{3.25}$$

has a polynomial solution if the accessory parameter is equal to zero and leading coefficient (2.25) satisfies the constraint

$$C_{1;1}(\varepsilon_{\pm,\sigma;m}\,|\,{}_1\mathbf{H};\pm,\sigma,\sigma)=-\,4m(\sigma\sqrt{|\,\varepsilon_{\pm,\sigma;m}\,|}\pm\lambda_o+m+1) \tag{3.26}$$

at the energy $\varepsilon=\varepsilon_{\pm,\sigma;m}$. [The latter constraint directly follows from (2.22) and (2.25), with n = 2m.]

Substituting (3,23) into (2.19) one finds

$$\Sigma_1[y;\bar{e},\bar{\rho}(\varepsilon\,|\,{}_1\mathbf{H};\pm,\sigma,\sigma)]=(1+\sigma\sqrt{|\,\varepsilon\,|})\left[{}^1\!/\!_2(1+\sigma\sqrt{|\,\varepsilon\,|})+(1\pm 2\lambda_o)\right]y. \tag{3.27}$$

and coupling the latter relation with (3.20) and (3.18′′), we come to the following explicit expression for polynomial (2.18):

$$C_1[y;\varepsilon\,|\,{}_1\mathbf{H};\pm,\sigma,\sigma]=\left[(1\pm\lambda_o+\sigma\sqrt{|\,\varepsilon\,|})^2-\mu_o^2+(1-b)\varepsilon\right]y \tag{3.28}$$

so that the accessory parameter in Heun equation (3.24) is indeed identically equal to 0. Constraint (3.25) thus turns into the quadratic equations

$$b\lambda^2_{1;\dagger_\pm,m}+2\lambda_{1;\dagger_\pm,m}(2m+1\pm\lambda_o)+(2m+1\pm\lambda_o)^2-\mu_o^2=0 \tag{3.29}$$

(cf. (5.1.29) and (6.1.19) in [11]) for energies $\varepsilon=\varepsilon_{\dagger_\pm,m}$ of AEH solutions of the CRSLE with Bose invariant (3.3). Here, in following [11, 21], we put

$$\lambda_{1;\mathbf{t}_\pm,m} \equiv \sigma_{\mathbf{t}_\pm} \sqrt{|\varepsilon_{\mathbf{t}_\pm,m}|}, \tag{3.31}$$

where $\mathbf{t}_+ = \mathbf{a}$ or $\mathbf{c}$, $\mathbf{t}_- = \mathbf{b}$ or $\mathbf{d}$, $\sigma_{\mathbf{t}} = +$ for $\mathbf{t} = \mathbf{b}$ or $\mathbf{c}$, and $\sigma_{\mathbf{t}} = -$ for $\mathbf{t} = \mathbf{a}$ or $\mathbf{d}$. It is essential that the appropriate solutions of the RCSLE with Bose invariant (3.6) also have the AEH form

$$\phi_{\mathbf{t}_\pm,m}[y\,|\,{}_1\mathcal{H}^{K\Im 1}_{\downarrow\mathbf{t}_\pm,m}] = y^{1\pm\lambda_o}\left(1-y^2\right)^{\frac{1}{2}\left(1+\lambda_{1;\mathbf{t}_\pm,m}\right)} \mathrm{Hp}_{2m}[y\,|\,{}_1\mathcal{H}^{K\Im 1}_{\downarrow\mathbf{t}_\pm,m};\mathbf{t}_\pm,m], \tag{3.32}$$

where

$$\mathrm{Hp}_{2m}[y\,|\,{}_1\mathcal{H}^{K\Im 1}_{\downarrow\mathbf{t}_+,m};\mathbf{t}_+,m] \equiv \mathrm{Hp}_{2m}[0\,|\,{}_1\mathcal{H}^{K\Im 1}_{\downarrow\mathbf{t}_+,m};\mathbf{t}_+,m]\mathrm{Hf}[y;\varepsilon_{\mathbf{t}_+,m}\,|+,\sigma_{\mathbf{t}_+},\sigma_{\mathbf{t}_+}] \tag{3.33a}$$

and

$$\mathrm{Hp}_{2m}[y\,|\,{}_1\mathcal{H}^{K\Im 1}_{\downarrow\mathbf{t}_-,m};\mathbf{t}_-,m] \equiv \mathrm{Hp}_{2m}[0\,|\,{}_1\mathcal{H}^{K\Im 1}_{\downarrow\mathbf{t}_-,m};\mathbf{t}_-,m]\mathrm{Hf}[1-y;\varepsilon_{\mathbf{t}_-,m}\,|-,\sigma_{\mathbf{t}_-},\sigma_{\mathbf{t}_-}], \tag{3.33b}$$

with the scale factor unambiguously determined by the condition that the leading coefficients of the polynomials in the right-hand side of (3.33a) and (3.33b) are equal to 1. Formally each solution (3.32) can be used to construct a rational SUSY partner $V[y(r)|{}_1^1\mathcal{H}^{K\Im 1}_{\mathbf{t},m}]$ of the radial potential $V[y(r)|{}_1\mathcal{H}^{K\Im 1}_{\downarrow\mathbf{t},m}]$. Since $V[y(r)|{}_1\mathcal{H}^{K\Im 1}_{\downarrow\mathbf{t},m}] = V[z(r)|{}_1\mathcal{G}^{K\Im 1}_{\downarrow\mathbf{t},m}]$ the potential constructed in such a way is nothing but the radial potential $V[z(r)\,|\,{}_1^1\mathcal{G}^{K\Im 1}_{\mathbf{t},m}]$ expressed in terms of y. In Section 6 we shall however present an example when the change of variable $z\to y$ does make a difference leading to a non-trivial set of polynomial solutions. It is explicitly confirmed below that

$$\mathrm{Hp}_{2m}[y\,|\,{}_1\mathcal{H}^{K\Im 1}_{\downarrow\mathbf{t},m};\mathbf{t},m] = \Pi_m[y^2;\bar{z}_{\mathbf{t},m}], \tag{3.34}$$

where $\bar{z}_{\mathbf{t},m}$ are zeros of the appropriate hypergeometric polynomial, so that the Heun polynomials in question have only simple zeros:

$$\mathrm{Hp}_{2m}[y \,|\, {}_1\mathbf{H}^{K\Im1}_{\downarrow\dagger,m}; \dagger, m] = \Pi_{2m}[y; \bar{y}_{\dagger,m}], \tag{3.34*}$$

where

$$y_{\dagger,m;m-j+1} = -\sqrt{z_{\dagger,m;j}}, \quad y_{\dagger,m;m+j} = \sqrt{z_{\dagger,m;j}} \quad \text{for } j=1, 2, \ldots, m \tag{3.35}$$

It is worth mentioning that bound energy states ($\dagger_+ = \mathbf{c}$) are described by the global AEH solutions

$$\phi_{\mathbf{c},v}[y \,|\, {}_1\mathbf{H}^{K\Im1}_{\downarrow\mathbf{c},v}] = y^{1+\lambda_o} \left(1-y^2\right)^{\frac{1}{2}\left(1+\lambda_{1;\mathbf{c},v}\right)} \Pi_{2v}[y; \bar{y}_{\mathbf{c},v}], \tag{3.36}$$

regular at all three singular points which implies that their polynomial components

$$S_{2v}^{(\lambda_o, \lambda_{1;\mathbf{c},v})}[y] \equiv \Pi_{2v}[y; \bar{y}_{\mathbf{c},v}], \tag{3.37}$$

belong to the class of Heine-Stieltjes polynomials [24, 50, 51]. We will refer to these particular representatives of Heun polynomials as Stieltjes-Van Vleck-Heun (SVH) polynomials to give a credit to Van Vleck [51] who studied the Heun reduction [17] of Heine-Stieltjes polynomials in great details.

Making use of the standard relation between general solutions of CRSLEs (3.2) and (3.7):

$$\Phi_{\bar{\sigma}}[y \,|\, {}_1\mathbf{H}^{K\Im1}] = \sqrt{\frac{dy}{dz}}\, \Phi_{\bar{\sigma}}[y^2 \,|\, {}_1\mathbf{G}^{K\Im1}] \tag{3.38}$$

and expressing the regular solutions of the former equation via hypergeometric functions [20]:

$$\Phi_{+,-,-}[y^2 \,|\, {}_1\mathbf{G}^{K\Im1}] = \sqrt{2}\, y^{1+\lambda_o} \left(1-y^2\right)^{\frac{1}{2}\left(1-\sqrt{|\varepsilon|}\right)} \times F[\alpha_{+,-,-}(\varepsilon;\lambda_o,\mu_o;a), \beta_{\pm,-,-}(\varepsilon;\lambda_o,\mu_o;a); 1+\lambda_o; y^2], \tag{3.39a}$$

and

$$\Phi_{-,+,+}[y^2 \mid {}_1\mathcal{G}^{K\Im 1}] = \sqrt{2}\, y^{1-\lambda_o} \left(1-y^2\right)^{\frac{1}{2}\left(1+\sqrt{|\varepsilon|}\right)} \qquad (3.39b)$$

$$\times\, F[\alpha_{-,+,+}(\varepsilon;\lambda_o,\mu_o;a), \beta_{-,+,+}(\varepsilon;\lambda_o,\mu_o;a); 1+\sqrt{|\varepsilon|}; 1-y^2],$$

where [20]

$$\alpha_{\pm,\sigma,\sigma}(\varepsilon;\lambda_o,\mu_o;a) + \beta_{\pm,\sigma,\sigma}(\varepsilon;\lambda_o,\mu_o;a) = 1 \pm \lambda_o + \sigma\sqrt{|\varepsilon|} \qquad (3.40)$$

and

$$\beta_{\pm,\sigma,\sigma}(\varepsilon;\lambda_o,\mu_o) - \alpha_{\pm,\sigma,\sigma}(\varepsilon;\lambda_o,\mu_o) = \sqrt{\mu_o^2 - a\varepsilon}. \qquad (3.40^*)$$

we come to the following explicit formulas for the Heun functions regular at one of the end points:

$$\mathrm{Hf}[y;\varepsilon \mid {}_1\mathbf{H}^{K\Im 1}; +,-,-] = F[\alpha_{+,--}(\varepsilon;\lambda_o,\mu_o;a), \beta_{+,--}(\varepsilon;\lambda_o,\mu_o;a); 1+\lambda_o; y^2]. \qquad (3.41a)$$

and

$$\mathrm{Hf}[y;\varepsilon \mid {}_1\mathbf{H}^{K\Im 1}; -,+,+] = F[\alpha_{-,+,+}(\varepsilon;\lambda_o,\mu_o;a), \beta_{-,+,+}(\varepsilon;\lambda_o,\mu_o;a); 1+\sqrt{-\varepsilon}; 1-y^2]. \qquad (3.41b)$$

Subtracting squares of (3.40) and (3.40*) and comparing the resultant expression with (3.28) we find

$$\alpha(\varepsilon \mid {}_1\mathbf{H}^{K\Im 1}; \pm,\sigma,\sigma)\beta(\varepsilon \mid {}_1\mathbf{H}^{K\Im 1}; \pm,\sigma,\sigma) = 4\alpha_{\pm,\sigma,\sigma}(\varepsilon;\lambda_o,\mu_o)\beta_{\pm,\sigma,\sigma}(\varepsilon;\lambda_o,\mu_o;a). \qquad (3.42)$$

Taking into account (3.40) one can also verify that

$$\alpha(\varepsilon \mid {}_1\mathbf{H}^{K\Im 1}; \pm,\sigma,\sigma) + \beta(\varepsilon \mid {}_1\mathbf{H}^{K\Im 1}; \pm,\sigma,\sigma) \qquad (3.42')$$

$$= 2[\alpha_{\pm,\sigma,\sigma}(\varepsilon;\lambda_o,\mu_o;a) + \beta_{\pm,\sigma,\sigma}(\varepsilon;\lambda_o,\mu_o;a)]$$

which implies that

$$\alpha(\varepsilon \mid {}_1\mathbf{H}^{K\Im 1}; \pm,\sigma,\sigma) = 2\alpha_{\pm,\sigma,\sigma}(\varepsilon;\lambda_o,\mu_o;a) \qquad (3.43a)$$

and

$$\beta(\varepsilon\,|\,{}_1\mathbf{H}^{K\Im 1};\pm,\sigma,\sigma) = 2\beta_{\pm,\sigma,\sigma}(\varepsilon;\lambda_o,\mu_o;a). \tag{3.43b}$$

We thus confirm that any Heun polynomial of order n = 2m satisfying of differential equations (3.25) coincides with a hypergeometric polynomial of order m in $y^2$:

$$Hp_{2m}[y\,|\,{}_1\mathbf{H}^{K\Im 1};\dagger] = \Pi_m[y^2;\bar{z}_{\dagger,m}] \tag{3.44}$$

$$= P_m^{(\lambda_o,\lambda_{1;\dagger,m})}(2y^2-1)/\tilde{k}_m(\lambda_o,\lambda_{1;\dagger,m}), \tag{3.44*}$$

where

$$\tilde{k}_m(\lambda,\mu) = \frac{(\lambda+\mu+m)_m}{m!}. \tag{3.45}$$

In particular the aforementioned SVH polynomials can be represented as

$$S_{2v}^{(\lambda_o,\lambda_{1;c,v})}[y] \equiv Hp_{2v}[y\,|\,{}_1\mathbf{H}^{K\Im 1};c]$$

$$= P_v^{(\lambda_o,\lambda_{1;c,v})}(2y^2-1)/\tilde{k}_v(\lambda_o,\lambda_{1;c,v}) \tag{3.46}$$

$$= \frac{P_v^{(\lambda_o,\lambda_{1;c,v})}(-1)}{\tilde{k}_v(\lambda_o,\lambda_{1;c,v})} F[-v,\beta_{+,\sigma,\sigma}(\varepsilon_{c,v};\lambda_o,\mu_o;a);1+\lambda_o;y^2]. \tag{3.46′}$$

Since the $v^{th}$-order hypergeometric polynomial in the right-hand side of (3.46′) has v positive roots between 0 and 1 the SVH polynomial in y has v pairs of roots of opposite sign with absolute values lying between 0 and 1, i.e. all the polynomial zeros lie between -1 and +1 as expected.

## 4.Basic SUSY ladder of h-PT potentials starting from the reflectionless sym-RM potential

About forty years ago the author [6-8] made an intriguing observation that the R@∞ solution of the Schrödinger equation with potential (1.1*) is expressible in terms of polynomials satisfying the Fuschian equation with 3 finite singular points. It was erroneously stated in [8] that all three singular points are regular. As a matter of fact,

Fuschian equation (32) in [8] has an irregular singular point at the origin. However, as already mentioned in Introduction, gauge transformation (1.3) does convert the cited equation into the Heun equation with polynomial solutions describing decayed-at-infinity wavefunctions for any arbitrarily selected energy $\varepsilon < 0$. For this reason we refer to (1.1*) as the PTpotential exactly solvable by Heun polynomials (Hp-ES). (Note that we distinguish between the terms 'exactly solvable' and 'exactly quantized' potentials.)

As initially pointed to by the author [5] R@∞ solutions of the Schrödinger equation with the reflectionless sym-RM potential $V_{0,t}(x)$are expressible via Jacobi polynomials so that the potential is also solvable by quadratures. We thus start our recurrent procedure from the R@∞ solution

$$\psi_{\mathbf{b}}^{(0,t+1)}(x;\kappa) \equiv ch^{t+1}x\left(-\frac{1}{ch\,x}\frac{d}{dx}\right)^{t+1} e^{-\kappa x} \tag{4.1}$$

$$= \kappa\, t!\; P_t^{(\kappa,-\kappa)}(th\,x)\, e^{-\kappa x} \text{ for } t= 0, 1, \ldots \tag{4.1′}$$

at an arbitrary energy $\varepsilon = -\kappa^2$. When the potential is restricted to the half-line solution (4.1′) for x>0 will be alternatively denoted as

$$\psi_{\mathbf{b}}^{(1,t)}(r;\kappa) \equiv \psi_{\mathbf{b}}^{(0,t+1)}(r;\kappa). \tag{4.2}$$

We then generate a sequence of isospectral potentials $V_{s'+1,t}(r)$ for s′= 1, …, s–1 via successive DTs

$$\psi_{\mathbf{b}}^{(s'+1,t)}(r;\kappa) = -\psi_{\mathbf{a}}^{(s',t)}(r)\frac{d}{dr}\frac{\psi_{\mathbf{b}}^{(s',t)}(r;\kappa)}{\psi_{\mathbf{a}}^{(s',t)}(r)} \qquad (s'= 1,\ldots, s-1) \tag{4.3}$$

with the R@O basic FFs defined via (8.1.2a) in [11]

$$\psi_{\mathbf{a}}^{(s',t)}(r) = sh^{s'} r\; ch^{t+s'} r \;\text{ for } s' = 1, 2, \ldots \tag{4.4}$$

which gives

$$\psi_{\mathbf{b}}^{(s,t)}(r;\kappa) \equiv sh^{s-1} r\; ch^{t+s-1} r\left(-\frac{2}{sinh(2r)}\frac{d}{dr}\right)^{s-1}\frac{\psi_{\mathbf{b}}^{(1,t)}(r;\kappa)}{ch^{t} r} \text{ fors} = 1,2, \ldots \tag{4.5}$$

Though multi-step DT (4.5) can be formally obtained from those [3, 4, 9] for trigonometric potential (1.1) by changing the argument x for *i*r the crucial difference however comes from the fact that DTs of our interest act on a R@∞ solution and therefore solution (4.6) must also decay at infinity. (One can easily verify that this is true for the DT of any potential with an exponential tail.) Another important characteristic of DTs (4.3) is that FF (4.4) is regular only at the origin whereas the FFs used by Darboux for trigonometric potential (4.1) are regular at both ends of the quantization interval and as a result erase the lowest eigenstate from an infinite set of bound energy levels.

As it has been discovered by the author [6, 7] the change of variable

$$y = th\, r \tag{4.6}$$

allows one to express elementary solutions (4.6) in terms of polynomials $P^{s,t}_{n'}[y;\kappa]$ of order n′ = t+2s–1 defined via the following recurrence relations

$$(2t+2s-1)P^{s+1,t}_{t+2s+1}[y;\kappa] = \left[ y(y^2-1)\frac{d}{dy} + ty^2 + \kappa y + 2s \right] P^{s,t}_{t+2s-1}[y;\kappa]\,, \tag{4.7}$$

starting from the weighted Jacobi polynomial

$$P^{1,t}_{t+1}[y;\kappa] \equiv y\, P^{0,t}_{t-1}[y;\kappa] = \frac{t!}{(2t-1)!!}\, y\, P^{(\kappa,-\kappa)}_{t}(y)\,. \tag{4.8}$$

The scale in the right-side of (4.8) is chosen via the requirement that the leading coefficient of the resultant polynomial is equal to 1 which is then also true for any other polynomial in sequence (4.7). An analysis of the mentioned expression for solutions (4.5) in terms of polynomials $P^{s,t}_{n'}[y;\kappa]$ of order n′ = t+2s-1:

$$\psi^{s,t}_{\mathbf{b}}(r;\kappa) = th^{-s} r\; P^{s,t}_{t+2s-1}[th\, r;\kappa]\; exp(-\kappa\, r) \tag{4.9}$$

Immediately shows that the polynomials $P^{s,t}_{n'}[y;\kappa]$ must have a single zero at y=0 (keeping in mind that the lower characteristic exponent of the Schrodinger equation at the singular point r=0 is equal to 1–s). This explains why the Fuschian equation for these polynomials

$$\left[y^2(y^2-1)\frac{d^2}{dy^2}+2y[(1-y)y^2+\kappa y+s]\frac{d}{dy}-(2s+t-1)t\,y^2-2s(\kappa y+1)\right]P^{s,t}_{t+2s-1}[y;\kappa]=0 \tag{4.10}$$

has an irregular singularity at y=0. An analysis of the latter equation again confirms that the polynomials in question must have zero at the origin and therefore can be represented as

$$P^{s,n+2-2s}_{n+1}[y;\kappa]= y\,Hp_n[y;\kappa;s]\ \text{for}\ 2s\leq n+2, \tag{4.11}$$

where

$$n = t-2s-2 \tag{4.12}$$

$$= \mu_o-\lambda_o-1, \tag{4.12*}$$

taking into account that

$$s=\lambda_o+1/2,\ t=\mu_o-\lambda_o. \tag{4.13}$$

The polynomials $Hp_n[y;\kappa;s]$ of order n in the right-hand side of (4.11) satisfy the Heun equation

$$\left[y(y^2-1)\frac{d^2}{dy^2}+2B_2[y;\kappa;s]\frac{d}{dy}+C_1[y;\kappa;s,n]\right]Hp_n[y;\kappa;s]=0 \tag{4.14}$$

where

$$B_2[y;\kappa;s]=(2-s)y^2+\kappa y+s-1 \tag{4.15}$$

and

$$C_1[y;\kappa;s,n]=-n(t+1)y-2\kappa(s-1). \tag{4.16}$$

As expected the leading coefficient of polynomial (4.16) is proportional to n -- the necessary condition for existence of a Heun polynomial of the n$^{th}$-order.

At this point we can directly relate our old results to the more general analysis presented in previous Section. Representing the h-PT potentials as

$$V_{s,t}(r) \equiv V[y(r) \,|\, {}_1\mathbf{H}^{011}], \tag{4.17}$$

where

$$V[y \,|\, {}_1\mathbf{H}^{011}] = (1-y^2)\left[\frac{s(s-1)}{y^2} - (s+t)(s+t-1)\right], \tag{4.18}$$

Substituting

$$y'(r) = ch^{-2} r = 1 - y^2(r). \tag{4.19}$$

$\{y, r\} = -2$, and (4.23) into (2.6) and (2.7) we come to the RefPF

$${}_1I^{o}_{H}[y;\lambda_o,\mu_o] = \frac{t(2s+t-1)}{1-y^2} - \frac{s(s-1)}{y^2} + \frac{1}{(1-y^2)^2} \tag{4.20}$$

which turns into (3.11) after substituting s and t for $\lambda_o$ and $\mu_o$ via (4.13).

One can directly verify that Heun equation (4.13) is nothing but the differential equation for the Heun function $Hf_1[y;\varepsilon|-+-]$ expressed in terms of the parameters s and t, instead of $\lambda_o$ and $\mu_o$. To prove it first note that second-order polynomial (4.14) can be alternatively represented as

$$B_2[y;\kappa;s] = y(y^2-1)\left[\frac{1-s}{y} + \frac{1+\kappa}{2(y-1)} + \frac{1-\kappa}{2(y+1)}\right], \tag{4.21}$$

in agreement with the canonical characteristic exponents for the given Heun function:

$$\bar{\rho}(\varepsilon \,|\, {}_1\mathbf{H};-+-)] = \tfrac{1}{2}(1-2\lambda_o, 1+\sqrt{|\varepsilon|}, 1-\sqrt{|\varepsilon|}), \tag{4.22}$$

with $\kappa \equiv \sqrt{|\varepsilon|}$. (Remember that the exponent difference for the singularity at r=0 in RCSLE (3.7) is equal to $2\lambda_o$=2s–1.) Polynomial (2.19), with the canonical characteristic exponents determined by (4.28), takes the form

$$\Sigma_1[y;\bar{e},\bar{\rho}(\varepsilon\,|\,{}_1\mathbf{H}^{101};-+-)] = \tfrac{1}{2}(1+\varepsilon)y + (1-2\lambda_o)(y+\sqrt{|\varepsilon|}). \tag{4.23}$$

By analogy with (3.28) we thus find that

$$C_1[y;-\kappa^2\,|\,{}_1\mathbf{H}^{101};-+-] = (\lambda_o^2-\mu_o^2+1-2\lambda_o)y + (1-2\lambda_o)\sqrt{|\varepsilon|}, \tag{4.24}$$

in agreement with (4.15). Substituting (4.28) into (2.24) shows that the coefficient $\beta_n(\bar{\rho})$ in eigenproblem (2.23) is independent of κ:

$$\beta_n(\bar{\rho}) = n-2s+3 \equiv t+1, \tag{4.25}$$

in agreement with (4.15).If s is a positive integer smaller than n + 2 then the infinite chain of eigenproblems(2.23) with n = 2s–1, 2s, …have a common nonnegative eigenvalue

$$q_1^{(n)} = 2\kappa(s-1) \geq 0$$

(4.26)

for any n >2s–2 and the corresponding LW polynomials form an infinite sequence of the Heun polynomials $\mathrm{Hp}_n[\varsigma;\kappa;s]$. Matrix elements (2.23a)-(2.23c) for the LW eigenproblem in question thus take the form

$$h_{j,j+1}^{(n)} = (j+1)(\lambda_o - j - 1), \tag{4.27a}$$

$$h_{j,j}^{(n)} = 2j\kappa, \tag{4.27b}$$

$$h_{j,j-1}^{(n)} = (j-n-1)(t+j). \tag{4.27c}$$

Substituting the matrix elements

$$h_{n,n}^{(n)} = 2n\kappa \tag{4.28}$$

and

$$h_{n,n-1}^{(n)} = -2(t+s-1), \tag{4.28′}$$

together with

$$G_n^{(n,1)}(\kappa; s, n) = 1, \tag{4.29}$$

into truncation condition (2.22′) then gives

$$G_{n-1}^{(n,1)}(\kappa; s, n) = \kappa. \tag{4.29′}$$

The next coefficient

$$G_{n-2}^{(n,1)}(\kappa; s, n) = \frac{(n-s)\kappa^2 - \tfrac{1}{2}n(n-2s+1)}{2n-2s+1} \tag{4.29″}$$

is then determined by substituting the matrix elements

$$h_{n-1,n}^{(n)} = -n(t-1),\ h_{n-1,n-1}^{(n)} = -4(n-1)\kappa,\ h_{n-1,n}^{(n)} = -n(t-1), \tag{4.30}$$

together with (4.29) and (4.29′), into (2.22) for j = n–1.

Representing (4.9) as

$$\psi_{\mathbf{b}}^{s,n-2s+2}[y;\kappa] = \varsigma^{1-s}\ Hp_n[y;\kappa;s]\left(\frac{1-y}{1+y}\right)^{\tfrac{1}{2}\kappa} \tag{4.31}$$

we come to the following explicit expression

$$Hf[y;\kappa;t,s;-,+,-] = Hp_n[y;\kappa;s]/\,Hp_n[1;\kappa;s], \tag{4.32}$$

for the Heun function

$$Hf[y;\kappa;t,s;-,+,-] = (1+y)^{\kappa}\ Hf[y;\kappa;t,s;-,+,+], \tag{4.33}$$

where we set

$$Hf[y;\kappa;t,s;-,+,+] \equiv Hf[y;-\kappa^2\,|\,{}_1H^{101};-,+,+]. \tag{4.34}$$

By determining the parameters $\alpha_{-,+,+}$ and $\beta_{-,+,+}$ of the hypergeometric function in the right-hand side of (3.39b) from (A.1) and (A.1*) in Appendix we can alternatively portray (4.31) as

$$\mathrm{Hf}[y;\kappa;t,s;-,+,-] = (1+y)^{\kappa}\, F[\tfrac{1}{2}(\kappa-n), \tfrac{1}{2}(\kappa-t-1); \kappa+1; 1-y^2]. \qquad (4.34')$$

Taking into account that the pair of quadratic equations (3.29) for b=1 turns into the following linear formulas:

$$\lambda_{1;\mathbf{a},m} \equiv -\kappa^{s,t}_{\mathbf{a},m} - t - 2s - 2m \qquad (4.35a)$$

$$\lambda_{1;\mathbf{a}',m} \equiv -\kappa^{s,t}_{\mathbf{a}',m} = t - 2m - 1 < 0, \qquad (4.35a')$$

$$\lambda_{1;\mathbf{b},m} \equiv \kappa^{s,t}_{\mathbf{b},m} = t + 2(s - m - 1) > 0, \qquad (4.35b)$$

$$\lambda_{1;\mathbf{b}',m} \equiv \kappa^{s,t}_{\mathbf{b}',m} = t - 2m - 1 > 0, \qquad (4.35b')$$

$$\lambda_{1;\mathbf{c},v} \equiv \kappa^{s,t}_{\mathbf{c},v} = t - 2v - 1 > 0 \quad \text{for } v = 0, 1, \ldots, v_{max}, \qquad (4.35c)$$

$$\lambda_{1;\mathbf{d},m} \equiv -\kappa^{s,t}_{\mathbf{d},m} = -t - 2m - 1 < 0, \qquad (4.35d)$$

$$\lambda_{1;\mathbf{d}',m} \equiv -\kappa^{s,t}_{\mathbf{d}',m} = t + 2(s - m - 1) < 0, \qquad (4.35d')$$

we can directly relate the Heun polynomials $\mathrm{Hp}_n[y;\kappa^{s,t}_{\mathbf{c},v};s]$ and $\mathrm{Hp}_n[y;\kappa^{s,t}_{\mathbf{b},m};s]$ to the weighted Jacobi polynomials in $2y^2 - 1$:

$$\tilde{k}_m(s+\tfrac{1}{2}, t-2v-1)\,\mathrm{Hp}_n(y; t-2v-1; s) = y^{2s-1}(y+1)^{t-2v-1} \times P_v^{(s+\frac{1}{2}, t-2v-1)}(2y^2-1) \qquad (4.36)$$

and

$$\tilde{k}_m(\tfrac{1}{2}-s, n-2m)\,\mathrm{Hp}_n(y; n-2m; s) = (y+1)^{n-2m}\, P_m^{(\frac{1}{2}-s, n-2m)}(2y^2-1), \qquad (4.36^*)$$

where the leading coefficient of the polynomial $P_m^{(\lambda,\mu)}(2y^2-1)$ in y is determined by (3.45). (Since t is assumed to be a positive integer the Hp-ES h-PT potential in question has at least one bound energy level and therefore AEH solutions $\mathbf{b}'$,m do not exist [21] in this particular case.)

To show that the derived formulas for coefficients (4.29′) and (4.29″) are consistent with relations (4.36) and (4.36*) between the Heun and Jacobi polynomials at $\kappa = \kappa^{s,t}_{\mathbf{c},v}$ and $\kappa = \kappa^{s,t}_{\mathbf{b},m}$, respectively, it is convenient to represent the latter relations in a slightly different form

$$\mathrm{Hp}_n(y; t-2v-1; s) = y^{2s-1}(y+1)^{t-2v-1}\Pi_v[y^2; \bar{z}_{\mathbf{c},v}] \tag{4.37}$$

and

$$\mathrm{Hp}_n(y; n-2m; s) = (y+1)^{n-2m}\ \Pi_m[y^2; \bar{z}_{\mathbf{b},m}], \tag{4.37*}$$

where the monomial product

$$\Pi_{2m}[y; \bar{y}_{\mathbf{t},m}] = \Pi_m[y^2; \bar{z}_{\mathbf{t},m}] \tag{4.38}$$

is nothing but the scaled hypergeometric polynomial in $y^2$ so that

$$\Pi_m[z; \bar{z}_{\mathbf{t},m}] = z^m + \frac{m(\sigma_{0;\mathbf{t}}\lambda_o + m)}{(\alpha_{\mathbf{t},m} + m - 1)(\beta_{\mathbf{t},m} + m - 1)} z^{m-1} + \ldots + \Pi_m[0; \bar{z}_{\mathbf{t},m}] \tag{4.39}$$

where $\alpha_{\mathbf{t},m} = -m$ and $\beta_{\mathbf{t},m} = \mu_o - m$, and

$$\Pi_m[0; \bar{z}_{\mathbf{t},m}] = \frac{(\alpha_{\mathbf{t},m} - 1)_m(\beta_{\mathbf{t},m} - 1)_m}{m!(\sigma_{0;\mathbf{t}}\lambda_o - 1)_m}. \tag{4.40}$$

We thus conclude that

$$\lim_{z\to\infty}\left(z^{1-m}\Pi_m[z; \bar{z}_{\mathbf{t},m}] - z\right) = -\frac{m(\sigma_{0;\mathbf{t}}\lambda_o + m)}{\mu_o - 1}. \tag{4.41}$$

By representing (4.29″) as

$$G^{(n,1)}_{n-2}(\kappa; s, n) = \tfrac{1}{2}(\kappa^2 - \kappa) + (n - \kappa)\frac{\kappa + 2\lambda_o - n}{4(\mu_o - 1)} \tag{4.41}$$

one can then directly verify that the coefficients of three largest powers of y in both sides of (4.37) and similarly (4.37*) do coincide as required.

For

$$0 < n - 2s - 2v - 1 < \kappa < n - 2s - 2v + 1 \tag{4.42}$$

the Heun polynomial $Hp_n[y;\kappa;s]$ has exactly v+1 zeros lying between 0 and 1 which implies that the zero of (2s–1)-th order at the origin splits into a simple positive zero and up to 2(s–1) negative zeros (if all of them are simple) for R@∞ solutions lying slightly above the $v^{th}$ bound energy level. As the energy increases and start approaching the next bound energy level up to 2s–1 negative zeros (if all of them are simple) move toward the origin to create the zero of the (2s–1)-th order at the energy $\varepsilon_v$.

It is more difficult to predict behavior of zeros near the outer singular point except that there must be a cluster of zeros near the point $\varsigma = 0$. This cluster expands by one-by-one as the monotonically increasing energy crosses the bound energy level.

## 5. Four basic recurrence formulas for Heun polynomials $Hp_n[y;\kappa;s]$

The Bose invariant $I[y;\varepsilon|{}_1\mathbf{H}^{101}]$ retains its form under four basic CLDTs with the FFs

$$\phi_{\mathbf{t};0}[y\,|\,{}_1\mathbf{H}^{101}] = y^{\frac{1}{2}+\sigma_{0;\mathbf{t}}\lambda_o}\left(1-y^2\right)^{\frac{1}{2}\lambda_{1;\mathbf{t},0}} \quad (\mathbf{t} = \mathbf{a}, \mathbf{b}, \mathbf{c}, \text{ or } \mathbf{d}), \tag{5.1}$$

where

$$\lambda_{1;\mathbf{a},0} \equiv -\kappa^{s,t}_{\mathbf{a},0} = -t - 2s \tag{5.2a}$$

$$= -n-2, \tag{5.2a$\dagger$}$$

$$\lambda_{1;\mathbf{b},0} \equiv \kappa^{s,t}_{\mathbf{b},0} = t + 2s - 2 \tag{5.2b}$$

$$= n, \tag{5.2b$\dagger$}$$

$$\lambda_{1;\mathbf{c},0} \equiv \kappa^{s,t}_{\mathbf{c},0} = t - 1 > 0, \tag{5.2c}$$

$$\lambda_{1;\mathbf{d},0} \equiv -\kappa^{s,t}_{\mathbf{d},0} = -t - 1 < 0. \tag{5.2d}$$

As it has been discussed in detail in [11] the DTs with the FF **b**,0 and **d**,0 decreases s by 1 when applied in the limit-point region $s > \frac{3}{2}$. On the contrary, The DTs with the FF **a**,0 and **c**,0 increases a value of this parameter by 1 for any s > 0. The DTs with both FF **a**,0 and **b**,0 do not change t thus keeping all the discrete energy levels unchanged (as anticipated [6] ). The DT with the FF **c**,0 (or **d**,0) decreases (increases) t by 2 and therefore erases (inserts) the ground energy level [6].

Since the potential in question has an exponential tail any DT converts the given R@∞ solution into the R@∞ solution of the resultant Schrödinger equation [11]. Substituting (4.29) into recurrence formula (A.9a) – (A.9d) in Appendix gives

$$\left[y(1-y^2)\frac{d}{dy}-(t+1)y^2-\kappa y-2s+1\right]Hp_n[y;\kappa;s] \\ =-(\kappa+n+2)\,f_{n,s;n+2,s+1}Hp_{n+2}[y;\kappa;s+1], \tag{5.4a}$$

$$\left[(1-y^2)\frac{d}{dy}+n\varsigma-\kappa\right]Hp_n[y;\kappa;s]=(n-\kappa)\,f_{n,s;n-2,s-1}\,y\,Hp_{n-2}[y;\kappa;s-1], \tag{5.4b}$$

$$\left[y(1-y^2)\frac{d}{d\varsigma}+ny^2-\kappa y-2s+1\right]Hp_n[y;\kappa;s] \\ =(t-1-\kappa)\,f_{n,s;n,s+1}Hp_n[y;\kappa;s+1], \tag{5.4c}$$

$$\left[(1-y^2)\frac{d}{dy}-(t+1)y-\kappa\right]Hp_n[y;\kappa;s]=-(\kappa+t+1)\,f_{n,s;n,s-1}\,yHp_n[y;\kappa;s-1], \tag{5.4d}$$

where

$$f_{n,s;n',s'}\equiv\frac{Hp_n[1;\kappa;s]}{Hp_{n'}[1;\kappa;s']}. \tag{5.5}$$

The first of the four formulas is nothing but derived earlier [6] recurrence relation (4.7) for the polynomials $P_{n'}^{s,t}[y;\kappa]$ after the latter are substituted for the Heun polynomials $Hp_n[y;\kappa;s]$ via (4.11). Comparing the leading coefficients of the polynomials in the left- and right sides of (5.4a) gives

$$f_{n,s;n+2,s+1}=\frac{Hp_n[1;\kappa;s]}{Hp_{n+2}[1;\kappa;s+1]}=\frac{2n-2s+3}{\kappa+n+2}, \tag{5.6a}$$

in agreement with (4.7). It then directly follows from this relation that

$$f_{n,s;n-2,s-1} = f^{-1}_{n-2,s-1;n,s} = \frac{\kappa + n}{2n - 2s + 1}. \tag{5.6b}$$

Recurrence formula (5.4a) in question is most important of the four because it allows one to construct the ladder of the Heun polynomials starting from scaled Jacobi polynomial

$$Hp_t[y;\kappa;1] = \frac{t!}{(2t-1)!!} P_t^{(\kappa,-\kappa)}(y) \tag{5.7}$$

and then increasing the polynomial order by 2 while keeping the parameter t unchanged. In particular, keeping in mind that

$$Hp_t[1;\kappa;1] = \frac{t!}{(2t-1)!!} P_t^{(\kappa,-\kappa)}(1) = \frac{(\kappa)_t}{(2t-1)!!} \tag{5.8}$$

we find

$$Hp_{t+2s-2}[1;\kappa;s] = \frac{(\kappa)_t}{(2t+2s-3)!!} \prod_{s'=1}^{s-1} (\kappa + t + 2s') \tag{5.9}$$

for s=1, 2,… and fixed t = n–2s+2 or alternatively

$$Hp_n[1;\kappa;s] = \prod_{j=1}^{s-1} (\kappa + n - 2j + 2) \frac{(\kappa)_{n-2s+2}}{(2n-2s+1)!!}. \tag{5.9′}$$

The latter relation allows one to compute the coefficients

$$f_{n,s;n,s+1} = \frac{\kappa + n - 2s + 1}{2n - 2s + 1} \tag{5.10}$$

and

$$f_{n,s;n,s-1} = \frac{n + t + 1}{\kappa + t + 1} \tag{5.10*}$$

appearing in the right-hand sides of (5.4c) and (5.4d).

After substituting (5.6a), (5.6b), (5.10), and (5.10*) into ladder relations (5.4a), (5.4b), (5.4c), and (5.4d), respectively, the latter take the form

$$\left[ y(1-y^2)\frac{d}{dy} - (n-2s+3)y^2 - \kappa y + 1 - 2s \right] Hp_n[y;\kappa;s] \tag{5.11a}$$

$$= -(2n-2s+3)Hp_{n+2}[y;\kappa;s+1],$$

$$\left[(1-y^2)\frac{d}{dy}+n\varsigma-\kappa\right]Hp_n[y;\kappa;s]=\frac{n^2-\kappa^2}{2n-2s+1}yHp_{n-2}[y;\kappa;s-1], \tag{5.11b}$$

$$\left[y(1-y^2)\frac{d}{dy}+ny^2-\kappa y-2s+1\right]Hp_n[y;\kappa;s]$$
$$=\frac{(n-2s+1)^2-\kappa^2}{2n-2s+1}Hp_n[\varsigma;\kappa;s+1], \tag{5.11c}$$

$$\left[(1-y^2)\frac{d}{dy}-(t+1)y-\kappa\right]Hp_n[y;\kappa;s]=-(n+t+1)\ yHp_n[y;\kappa;s-1] \tag{5.11d}$$

which constitute the main result of this section. Making use of (4.29″) and (4.2″) one can directly verify that the leading coefficients of the polynomial in both sides of recurrence formulas (5.11b) and (5.11c) coincide despite the fact that the polynomial expression in the left sides formally have larger orders.

## 6. Rational SUSY partners of the h-PT potential exactly solvable by Heine polynomials

As demonstrated in Section 3 the radial *r*-GRef potential $V[z\,(r)|_1\mathcal{G}^{K\Im 1}]$ can be represented in an alternative form such that the appropriate Schrödinger equation expressed in terms of new variable (3.1) is exactly solvable by Heun functions in ς. At the first glance this may not look like a significant development. Why would one need to express the solutions in terms of rather complicated Heun functions when the Schrödinger equation is exactly solved by hypergeometric functions with the well-known asymptotics? The example presented in Section 4 provides an insensitive for this new turn in the conventional theory of rational potentials exactly solvable via hypergeometric functions. The crucial point is that, as clarified below, the R@∞ solution of CSLE (3.7) has an AEH form

$$\phi_{\mathbf{b}}[y;\kappa;s,n]=y^{1-s}(1-y)^{(1+\kappa)/2}(1+y)^{(1-\kappa)/2}Hp_n[y;\kappa;s], \tag{6.1}$$

at an arbitrary energy $\varepsilon=-\kappa^2$. As a result, the CLDT using AEH FF (6.1), where $\kappa = \kappa_1 > \kappa_{\mathbf{c}}^{s,t}$ , generates a rational SUSY partner $^1V\,[y;\kappa_1;s,n]$ exactly solvable by polynomials.

One can easily obtain (6.1) from (4.9) and (4.11) by taking into account that the wavefunction is related to the solution of the CSLE via (2.29), where the density function in the particular case of the h-PT potential is given by

$$ {}_1\wp_H[y;0,1] = \frac{1}{(1-y^2)^2}. \tag{6.2} $$

Making use of Suzko's reciprocal formula (2.36) and assuming that all the zeros $y_j(\kappa_1;s,n)$ of the Heun polynomial in the right-hand side of (6.1) are simple one can represent the FF for the inverse CLDT as

$$ {}^{\star}\phi_{\mathbf{b}}[y;\kappa_1;s,n] = y^{s-1}(1-y)^{(1-\kappa_1)/2}(1+y)^{(1+\kappa_1)/2}\Pi_n^{-1}[y;\bar{y}(\kappa_1;s,n)] \tag{6.3} $$

which is nothing but a basic AEH solution of the partner CSLE with the Bose invariant

$$ I^o[y;\varepsilon;{}^{\star}e(s,n\,|\,\kappa_1);s,n\,|\,\kappa_1] = I^o[y;{}^{\star}e(s,n\,|\,\kappa_1);s,n\,|\,\kappa_1] + {}_1\wp_H[y;0,1]\,\varepsilon, \tag{6.4} $$

where

$$ {}^{\star}e(s,n\,|\,\kappa_1) \equiv [0,1,-1,\bar{y}(\kappa_1;s,n)]. \tag{6.5} $$

By decomposing density function (6.2) as

$$ {}_1\wp_H[y;0,1] = \frac{1}{2(1-y)^2} + \frac{1}{2(1+y)^2} + \frac{1}{2(1-y^2)} \tag{6.6} $$

we come to the following explicit expression for the reference PF in question

$$ I^o[y;{}^{\star}e(\kappa_1;s,n);s,n\,|\,\kappa_1] = -\frac{(s-1)(s-2)}{y^2} + \frac{1+y^2}{(1-y^2)^2} - \sum_{j=1}^{n} \frac{2}{[y-y_j(\kappa_1;s,n)]^2} + \frac{{}^{\star}O_{n+1}^o[y;s,n\,|\,\kappa_1]}{4y(y^2-1)\Pi_n[y;\bar{y}(\kappa_1;s,n)]}\,, \tag{6.7} $$

where

$$ {}^{*}O^{o}_{n+1}\left[y;\kappa;s,n\right] = -4\Sigma_{n+1}[y;{}^{*}\bar{e}(\kappa;s,n),{}^{*}\bar{\rho}(\kappa;s,n)] - \kappa^{2}\; y\,\Pi_{n}[y;\bar{y}(\kappa;s,n)], $$

(6.8)

with the vector parameters $\bar{e}$ and $\bar{\rho}$ of polynomial (2.18) set to (6.5) and to

$$ {}^{*}\bar{\rho}(\kappa_1;s,n) = [s-1, \tfrac{1}{2} - \tfrac{1}{2}\kappa_1, \tfrac{1}{2} + \tfrac{1}{2}\kappa_1, -\bar{1}_n], \qquad (6.9) $$

respectively. (Here $\bar{1}_n$ stands for the n-element row formed by ones.) Note that the CLDT keeps the residues of the appropriate second-order poles unchanged since the TP associated with the density function (6.2) does not vanish at the singular points y=±1 [32].

The assumption that all the zeros $\varsigma_j(\kappa_1;s,n)$ of the Heun polynomial $Hp_n[y;\kappa;s]$ are simple would require a more cautious analysis in the future since it is definitely invalid for polynomial (4.33*) at any of the factorization energies $\varepsilon_{\mathbf{b},m}$ ( $\kappa_1 = \kappa^{s,t}_{\mathbf{b},m}$ ).

Making use of the generic expression

$$ W\{f[y]\psi_1[y], f[y]\psi_2[y]\} = f^{2}[y]W\{\psi_1[y],\psi_2[y]\} \qquad (6.10) $$

for the Wroskian of two function and choosing

$$ f[y] = \sqrt{1-y^{2}}\;{}_1\wp^{1/4}_{H}[y;0,1]\; y^{1-s} \qquad (6.11) $$

we can represent the R@∞ solution of the RCSLE with Bose invariant (6.4) as

$$ \phi_{\mathbf{b}}[y;\kappa;s,n\,|\,\kappa_1] = (1-y^{2})W\{\phi[y;\kappa_1;s,n],\phi_{\mathbf{b}}[y;\kappa;s,n]\}/\phi_{\mathbf{b}}[y;\kappa_1;s,n] \qquad (6.12) $$

$$ = y^{1-s}(1-y)^{(1+\kappa)/2}(1+y)^{(1-\kappa)/2}\,\frac{{}^{1}P_{2n+1}[y;\kappa;s,n\,|\,\kappa_1]}{\Pi_n[y;\bar{y}(\kappa_1;s,n)]} $$

$$ \text{for}\;\; \kappa \neq \kappa^{s,n+2-2s}_{\mathbf{c},v} \quad (v = 0, 1,\dots,v_{max}), \qquad (6.12^*) $$

where $v_{max}$ is defined via inequality (4.32c). Since the characteristic exponent at the origin is equal to 2–s for any solution irregular at the given singular point the polynomial determinant (PD)

$$ {}^{1}P_{2n+1}[y;\kappa;s,n\,|\,\kappa_1] \equiv \begin{vmatrix} Hp_n[y;\kappa_1;s] & Hp_n[y;\kappa;s] \\ P_{n+1}[y;\kappa_1;s] & P_{n+1}[y;\kappa;s] \end{vmatrix}, \tag{6.13} $$

where

$$ P_{n+1}[y;\kappa;s] \equiv \left[(1-y^2)\frac{d}{dy} - \kappa\right] Hp_n[y;\kappa;s], \tag{6.14} $$

must vanish at y = 0 which implies that the polynomial

$$ Hi_{2n}[y;\kappa;s,n\,|\,\kappa_1] \equiv \frac{{}^{1}P_{2n+1}[y;\kappa;s,n\,|\,\kappa_1]}{(\kappa_1-\kappa)y}, \tag{6.15} $$

where

$$ {}^{1}P_{2n+1}[y;\kappa;s,n\,|\,\kappa_1] = (1-y^2)W\{Hp_n[y;\kappa_1;s], Hp_n[y;\kappa;s]\} \tag{6.16} $$

$$ +\,(\kappa_1-\kappa)Hp_n[y;\kappa_1;s]Hp_n[y;\kappa;s], $$

is a solution of the Heine equation. One can directly confirm (6.15) by analyzing Heun equation (4.13) which shows that

$$ \dot{H}p_n[0;\kappa;s] = \kappa\, Hp_n[0;\kappa;s] \tag{6.17} $$

for any $\kappa$ (including $\kappa = \kappa_1$), with dot used to denote the first derivative with respect y. As mentioned introduction we refer to polynomial solutions (6.15) as 'Heun seed' (HpS) Heine polynomials by analogy with the term 'Gauss seed' (GS) Heine polynomials' introduced in our analysis of P-CEQ SUSY partners of r-GRef potentials. The outlined procedure thus allows one to construct a rational potential exactly solvable via elementary functions for each pair of positive integers t and s and any value of the parameter $\kappa_1$ larger than $\kappa^{s,t}_{\mathbf{c},0}$.

Substituting (4.37) into (6.16) evaluated at $\kappa = \kappa^{s,t}_{\mathbf{c},v}$ one finds

$$ {}^{1}P_{2n+1}[y;\kappa^{s,t}_{\mathbf{c},v};s,n\,|\,\kappa_1] = y^{2s-2}\,(y+1)^{\kappa^{s,t}_{\mathbf{c},v}}\;{}^{1}P_{n+2v}[y;s,n\,|\,\kappa_1;\mathbf{c},v], \qquad (6.18) $$

As proven below the order of the polynomial

$$ \begin{aligned} {}^{1}P_{n+2v}[y;s,n\,|\,\kappa_1;\mathbf{c},v] &\equiv y(1-y^2)W\{Hp_n[y;\kappa_1;s],\Pi_{2v}[y;\bar{y}_{\mathbf{c},v}]\} \\ &+[(2s-1)(1-y^2)-\kappa^{s,t}_{\mathbf{c},v}y^2+\kappa_1 y]Hp_n[y;\kappa_1;s]\Pi_{2v}[y;\bar{y}_{\mathbf{c},v}] \end{aligned} \qquad (6.19) $$

is equal to n+2v but not to n+2m+2 as one could formally expect from its definition (6.19). By combining (6.12*) and (6.18) the $v^{th}$ eigenfunction of the RCSLE with Bose invariant (6.4) can be thus represented as

$$ \phi_{\mathbf{c},v}[y;s,n\,|\,\kappa_1] = y^{s-1}(1-y^2)^{\frac{1}{2}(1+\kappa^{s,t}_{\mathbf{c},v})}\,\frac{{}^{1}P_{n+2v}[y;s,n\,|\,\kappa_1;\mathbf{c},v]}{\Pi_n[y;\bar{y}]}. \qquad (6.20) $$

Since the power of y in the right-hand side of this relation coincides with the characteristic exponent of the R@O solution of the given CSLE polynomial (6.19)must remain finite as $y\to 0$ and therefore it is proportional to the quantized HpS Heine polynomial (HpS q-HiP)

$$ {}^{1}P_{n+2v}[y;s,n\,|\,\kappa_1;\mathbf{c},v] = (n+1-2s-2v)Hi_{n+2v}[y;s,n\,|\,\kappa_1;\mathbf{c},v], \qquad (6.21) $$

Where the scale factor was determined based on the identities (6.14) and (6.19).To prove that the order of polynomial (6.21) is indeed equal to n+2v, let us first represent (6.19) and (6.21) as

$$ \upsilon_{\mathbf{c},v}(s,n\,|\,\kappa_1)Hi_{n+2v}[y;s,n\,|\,\kappa_1;\mathbf{c},v] = \hat{g}_v(s,n\,|\,\kappa_1)\Pi_{2v}[y;\bar{y}_{\mathbf{c},v}], \qquad (6.22) $$

where the operator

$$ \begin{aligned} \hat{g}_v(s,n\,|\,\kappa_1) &\equiv \Pi_{2v}[y;\bar{y}_{\mathbf{c},v}]\left[y(1-y^2)\frac{d}{dy}+2s-1+\kappa_1 y-(n-2v)y^2\right] \\ &-y(1-y^2)\dot{\Pi}_{2v}[y;\bar{y}_{\mathbf{c},v}] \end{aligned} \qquad (6.23) $$

is referred to below as the HpS q-HiP generator (q-HiPG). Making use of its alternative representation

$$\hat{g}_{v}(s,n\,|\,\kappa_1) = \Pi_n[y;\bar{y}(\kappa_1;s,n)]\left[y(1-y^2)\frac{d}{dy} + 2vy^2\right] + p_n[y;s,n\,|\,\kappa_1] \qquad (6.23^*)$$

coupled with (5.29), (5.29′, and (5.29″), shows that

$$p_n[y;s,n\,|\,\kappa_1] \equiv (2s-1+\kappa_1 y - n\,y^2)\Pi_n[y;\bar{y}(\kappa_1;s,n)] - y(1-y^2)\,\dot{\Pi}_n[y;\bar{y}(\kappa_1;s,n)], \qquad (6.24)$$

is the n-order polynomial with the leading coefficient

$$p_{n;n}(s,n\,|\,\kappa_1) = \frac{\kappa_1^2 - (n-2s+1)^2}{2n-2s+1} \qquad (6.25)$$

so that

$$\upsilon_{\mathbf{c},v}(s,n\,|\,\kappa_1) = \frac{2v(\mu_o - \lambda_o - 1)}{\mu_o - 1} + p_{n;n}(s,n\,|\,\kappa_1) \qquad (6.26)$$

$$= 2v + \frac{\kappa_1^2 - (n-2s+1)^2 - 2v(2s+2v-1)}{2n-2s+1}. \qquad (6.26')$$

According to (5.11c) and (5.11d), CLDTs with the FF **c**,0 and **d**,0 do not change the order of the Heun polynomial $Hp_n[\varsigma;\kappa;s]$ and therefore the potential ${}^1V_{\mathbf{b}}[y;\kappa_1;s,n]$ retains its form under the DT with the nodeless normalizable FF. Historically potentials retaining their form under DTs with FFs of this type are referred to as 'shape-invariant', in following the imprecise translation [54] of the term 'form-invariant' introduced by Gendenshtein in this context. This observation is the direct consequence of the following general rule [32, 11]

*If both rational Liouville potential and an AEH solution retain their analytical form under a CLDT with a basic FF then the resultant partner potential generated by*

*means of this solution is form-invariant under the second-step CLDT using both FFs as its seeds.*

In this context the proposition that the AEH solution in question retains its analytical form implies that the order of the polynomial forming this solution remains the same and as a result the number of singular points is unaffected by the mentioned second-step CLDT.

An analysis of recurrence formula (A.6c) in Appendix shows that the CLST with the FF $\mathbf{c}$,0 keeps unchanged parameter (A.3b) and therefore this should be also true for the order of the appropriate Jacobi polynomial if this parameter is set to a non-positive integer. We thus conclude that the rational SUSY partners $V[z\,|\,{}^{1}_{1}G^{101}_{\mathbf{b},m}]$ of the h-PT potential are 'shape-invariant' for any values of the parameters s and t larger than $3/2$ and 0, respectively. As already pointed in [11], we only need to consider the limit-point (LP) range of the parameter s since the condition

$$\kappa^{s,t}_{\mathbf{b},m} = t + 2(s - m - 1) > \kappa^{s,t}_{\mathbf{c},0} = t - 1 \tag{6.27}$$

holds only for $m < s - 1/2$. For any integer values of the parameters t > 0 and s > 1 the P-CEQ potential $V[z(r)\,|\,{}^{1}_{1}G^{101}_{\mathbf{b},m}]$ becomes also exactly solvable via elementary functions proportional to HpS HiPs in *th* r.

Representing the GS HiPG defined via (2.57) in [21]:

$${}_{1}\hat{g}(\Delta\lambda, \Delta\nu; \bar{z}_{\mathbf{b},m}) = \Pi_m[y^2; \bar{z}_{\mathbf{b},m}]\, {}_{1}\hat{g}(\Delta\lambda, \Delta\nu) - \tfrac{1}{2} y(1 - y^2)\, \dot{\Pi}_m[y^2; \bar{z}_{\mathbf{b},m}] \tag{6.28}$$

where

$${}_{1}\hat{g}(\Delta\lambda, \Delta\nu) \equiv z(1 - z)\frac{d}{dz} + \tfrac{1}{2}[\Delta\lambda(1 - z) - \Delta\nu\, z] \tag{6.29}$$

$$= \tfrac{1}{2}\left[\, y(1 - y)\frac{d}{dy} + \Delta\lambda(1 - y^2) - \Delta\nu\, y^2 \,\right], \tag{6.29*}$$

gives

$$\hat{g}_v(s,n\,|\,\kappa^{s,t}_{\mathbf{b},m}) = 2(y+1)^{n-2m}\,{}_1\hat{g}(2\lambda_o,\kappa^{s,t}_{\mathbf{c},v}-\kappa^{s,t}_{\mathbf{b},m};\bar{z}_{\mathbf{b},m}), \tag{6.31}$$

with

$$\kappa^{s,t}_{\mathbf{c},v}-\kappa^{s,t}_{\mathbf{b},m} = 1+2(m-v-s). \tag{6.32}$$

We thus conclude that

$$\mathrm{Hi}_{n+2v}[y;s,n\,|\,\kappa^{s,t}_{\mathbf{b},m};\mathbf{c},v] = (y+1)^{n-2m}\mathrm{Hi}_{m+v}[y^2\,|\,{}^{1}_{1}\mathcal{G}^{101}_{\mathbf{b},m};\mathbf{c},v] \tag{6.33}$$

where the GS q-HiP in the right-hand side is defined via (2.58) in [21]; namely,

$${}_1\upsilon_{\mathbf{b},m;\mathbf{c},v}\mathrm{Hi}_{m+v}[z\,|\,{}^{1}_{1}\mathcal{G}^{101}_{\mathbf{b},m};\mathbf{c},v]$$

$$= {}_1\hat{g}(2\lambda_o,\kappa^{s,t}_{\mathbf{c},v}-\kappa^{s,t}_{\mathbf{b},m};\bar{z}_{\mathbf{b},m})\Pi_v[z;\bar{z}_{\mathbf{c},v}] \tag{6.34}$$

$$= z(1-z)W\{\Pi_m[z;\bar{z}_{\mathbf{b},m}],\Pi_v[z;\bar{z}_{\mathbf{c},v}]\} \tag{6.34′}$$

$$+[\lambda_o-(m-v)z]\Pi_m[z;\bar{z}_{\mathbf{b},m}]\Pi_v[z;\bar{z}_{\mathbf{c},v}].$$

An analysis of the (m+v–1)- and (m+v–2)-order coefficients of the polynomial Wroskian in the right-hand side of (6.34′) shows that

$$z^{2-m-v}W\{\Pi_m[z;\bar{z}_{\mathbf{b},m}],\Pi_v[z;\bar{z}_{\mathbf{c},v}]\} = (v-m)z^{1-m-v}\Pi_m[z;\bar{z}_{\mathbf{b},m}]\Pi_v[z;\bar{z}_{\mathbf{c},v}]$$

$$+\frac{[v(\lambda_o+v)+m(m-\lambda_o)]}{\mu_o-1}+O(z^{-1}) \tag{6.35}$$

so that

$${}_1\upsilon_{\mathbf{b},m;\mathbf{c},v} = \frac{v(\mu_o-\lambda_o-v-1)+(\lambda_o-m)(\mu_o-m-1)}{\mu_o-1}. \tag{6.36}$$

On other hand, evaluation of $\kappa_1$–dependent scale factor (6.26′) at $\kappa_1=\kappa^{s,t}_{\mathbf{b},m}$ gives

$$\upsilon_{\mathbf{c},v}(s,n\,|\,\kappa^{s,t}_{\mathbf{b},m}) = \frac{(2n-2s+1-2m)(2s-2m-1)+2v(t-v-1)}{2n-2s+1}. \qquad (6.37)$$

Substituting the parameters s and t, instead of $\lambda_o$ and $\mu_o$, we find

$$\upsilon_{\mathbf{c},v}(s,n\,|\,\kappa^{s,t}_{\mathbf{b},m}) = 2\,{}_1\upsilon_{\mathbf{b},m;\mathbf{c},v}, \qquad (6.38)$$

as expected from (6.31). It is worth stressing again that the latter relation holds iff t and s are positive integers.

RCSLE (3.7) with K=1 and ℑ=0 also has two sequences of infinitely many nodeless R@O AEH solutions

$$\phi_{\mathbf{t}_+,m}[y;t,s] = y^{1+\lambda_o}\left(1-y^2\right)^{\frac{1}{2}\left(1+\lambda_{1;\mathbf{t}_+,m}\right)}\Pi_{2m}[y;\bar{y}_{\mathbf{t}_+,m}] \qquad (6.39)$$

($\mathbf{t}_+ = \mathbf{a}$ or $\mathbf{a}'$).

Here

$$\lambda_{1;\mathbf{a},m} = -t-2\lambda_o-2m-1<0 \qquad (6.40)$$

or

$$\lambda_{1;\mathbf{a}',m} = t-2m-1 < -\kappa^{s,t}_{\mathbf{c},0} = 1-t \;\; (m > t-1). \qquad (6.40')$$

Making again use of Suzko's reciprocal formula (2.36) and taking into account that all the zeros of the Jacobi polynomial are simple one can represent the FF for the inverse CLDT as

$$^{*}\phi_{\mathbf{t}_+,m}[y\,|\,{}^{1}_{1}\mathbf{H}^{101}_{\mathbf{t}_+,m}] = y^{-s}(1-y)^{\frac{1}{2}(1+\kappa_{\mathbf{t}_+,m})}(1+y)^{\frac{1}{2}(1-\kappa_{\mathbf{t}_+,m})}\Pi^{-1}_{2m}[y;\bar{y}_{\mathbf{t}_+,m}] \qquad (6.41)$$

which is a basic AEH solution of the CSLE with the Bose invariant

$$I[y;\varepsilon\,|\,{}^{1}_{1}\mathbf{H}^{101}_{\mathbf{t}_+,m}] = I^{o}[y;{}^{*}\bar{e}_{\mathbf{t}_+,m};s,t;\mathbf{t}_+,m] + {}_1\wp_H[y;0,1]\varepsilon, \qquad (6.42)$$

where

$$^{*}\bar{e}_{\mathbf{t}_+,m} \equiv (0,1,-1,\bar{y}_{\mathbf{t}_+,m}). \qquad (6.43)$$

Similarly to (6.5), the reference PF in the right-hand side of (6.41) is given by the relation

$$I^{o}[y;{}^{*}\bar{e}_{\dagger_{+},m};s,t;\dagger_{+},m] = -\frac{s(s+1)}{y^{2}} + \frac{1+y^{2}}{(1-y^{2})^{2}} - \sum_{j=1}^{2m}\frac{2}{[y-y_{\dagger_{+},m;j}]^{2}}$$

$$+\frac{{}^{*}O^{o}_{2m+1}\left[y;s,n;\dagger_{+},m\right]}{4y(y^{2}-1)\Pi_{2m}\left[y;\bar{y}_{\dagger_{+},m}\right]}, \tag{6.44}$$

where

$${}^{*}O^{o}_{m+1}\left[y;s,t\,|\,\dagger_{+},m\right] = -4\Sigma_{m+1}[y;{}^{*}\bar{e}_{\dagger_{+},m},{}^{*}\bar{\rho}_{\dagger_{+},m}]$$

$$+\varepsilon_{\dagger_{+},m}\,y\,\Pi_{2m}\left[y;\bar{y}_{\dagger_{+},m}\right], \tag{6.45}$$

with

$${}^{*}\bar{\rho}_{\dagger_{+},m} = (s+1, \tfrac{1}{2}+\tfrac{1}{2}\kappa, \tfrac{1}{2}+\tfrac{1}{2}\kappa, -\bar{1}_{n}). \tag{6.46}$$

Representing the R@∞ solution of the CSLE with Bose invariant (6.42) as

$$\phi_{\mathbf{b}}[y;\kappa;s,n;\dagger_{+},m] = (1-y^{2})W\{\phi_{\dagger_{+},m}[y;t,s],\phi_{\mathbf{b}}[y;\kappa;s,n]\}/\phi_{\dagger_{+},m}[y;t,s] \tag{6.47}$$

and making use of (6.10), with

$$f[y] = \sqrt{1-y^{2}}\;{}_{1}\wp_{H}^{1/4}[y;0,1] \tag{6.48}$$

one finds

$$\phi_{\mathbf{b}}[y;\kappa;s,n\,|\,\dagger_{+},m] = y^{-s}(1-y)^{(1+\kappa)/2}(1+y)^{(1-\kappa)/2}\frac{{}^{1}P_{n+2m+2}[y;\kappa;s,n\,|\,\dagger_{+},m]}{\Pi_{m}[y;\bar{y}_{\dagger_{+},m}]}$$

$$\dagger_{+} = \mathbf{a} \text{ or } \mathbf{a}'. \tag{6.49}$$

The polynomials in the second row of the PD

$${}^{1}P_{n+2m+2}[y;\kappa;s,n\,|\,\dagger_{+},m] \equiv \begin{vmatrix} \Pi_{2m}[y;\bar{y}_{\dagger_{+},m}] & Hp_{n}[y;\kappa;s] \\ 2G_{m+1}[y^{2};s,\lambda_{1;\dagger_{+},m}\,|\,\dagger_{+}] & H^{(s)}_{n+2}[y;\kappa] \end{vmatrix} \tag{6.50}$$

are defined as follows

$$G_{m+1}[z;\rho_0,\rho_1 \mid \mathbf{t}_+] \equiv {}_1\hat{g}(\rho_0,\rho_1)\Pi_m[z;\bar{z}_{\mathbf{t}_+,m}] \tag{6.51}$$

and

$$H^{(s)}_{n+2}[y;\kappa] \equiv yP_{n+1}[y;\kappa;s] + (1-s)(1-y^2)Hp_n[y;\kappa;s], \tag{6.51′}$$

with the leading coefficients equal to

$$G_{m+1;m+1}(\rho_0,\rho_1 \mid \mathbf{t}_+) \equiv -{}^1\!/_2(\rho_0+\rho_1) - m \tag{6.52}$$

and

$$H^{(s)}_{n+2;n+2}(\kappa) = s-1-n, \tag{6.52′}$$

respectively. As expected the leading coefficient of polynomial (6.50) vanishes if we choose $\mathbf{t}_+ = \mathbf{c}$:

$$^1P_{n+2v+2;n+2v+2}(\kappa;s,n \mid \mathbf{c},v) = 2s-1-n+\lambda_{1;\mathbf{c},0} = 0. \tag{6.53c}$$

Otherwise the polynomial has the negative leading coefficient

$$^1P_{n+2m+2;n+2m+2}(\kappa;s,n \mid \mathbf{t}_+,m) = -\kappa^{s,t}_{\mathbf{c},0} - \kappa^{s,t}_{\mathbf{t}_+,0} \quad (\mathbf{t}_+ = \mathbf{a} \text{ or } \mathbf{a}') \tag{6.53a}$$

independent of m. Since the power exponent of y in the right-hand side of (6.49) coincides with the lowest ChExp at the origin the PD must be proportional to the HpSHiP

$$^1P_{n+2m+2}[y;\kappa;s,n \mid \mathbf{t}_+,m] = -(\kappa^{s,t}_{\mathbf{t}_+,0} + t - 1)\,Hi_{n+2m+2}[y;\kappa;s,n \mid \mathbf{t}_+,m]. \tag{6.54}$$

Substituting (4.37) into (6.51′) at $\kappa = \kappa^{s,t}_{\mathbf{c},v}$ gives

$$H^{(s)}_{n+2}[y;\kappa^{s,t}_{\mathbf{c},v}] = 2y^{2s-1}(y+1)^{\kappa^{s,t}_{\mathbf{c},v}}\,G_{v+1}[y^2;s,\kappa^{s,t}_{\mathbf{c},v} \mid \mathbf{c}] \tag{6.55}$$

so that

$$\mathrm{Hi}_{n+2m+2}[y;\kappa^{s,t}_{\mathbf{c},v};s,n\,|\,\mathbf{t}_+,m]=y^{2s-1}(y+1)^{t-2v-1} \qquad (6.56)$$

$$\times\mathrm{Hi}_{m-v+1}[y^2\,|\,{}^1_1\mathcal{G}^{101}_{\mathbf{t}_+,m};\mathbf{c},v],$$

where

$$\mathrm{Hi}_{m-v+1}[z\,|\,{}^1_1\mathcal{G}^{101}_{\mathbf{t}_+,m};\mathbf{c},v]=\mathrm{G}_{m+1}[z;s,-\kappa^{s,t}_{\mathbf{t}_+,m}\,|\,\mathbf{t}_+]\Pi_v[z;\bar{z}_{\mathbf{c},v}] \qquad (6.57)$$

$$-\mathrm{G}_{v+1}[z;s,\kappa^{s,t}_{\mathbf{c},v}\,|\,\mathbf{c}]\Pi_m[z^2;\bar{z}_{\mathbf{t}_+,m}].$$

Taking into account that

$$\mathrm{Hi}_{m+1}[z\,|\,{}^1_1\mathcal{G}^{101}_{\mathbf{t}_+,m};\mathbf{c},0]=\mathrm{Hi}_{m+1}[z\,|\,{}^1_1\mathcal{G}^{101}_{\mathbf{c},0};\mathbf{t}_+,m] \qquad (6.58)$$

$$=P^{(\lambda_o,\lambda_{1;\mathbf{t}_+,m})}_{m+1}(2z-1)/\,\square k_m(\lambda_o,\lambda_{1;\mathbf{t}_+,m}) \qquad (6.58^*)$$

we conclude that, contrary to the SUSY partners $V[z\,|\,{}^1_1\mathcal{G}^{101}_{\mathbf{b},m}]$, the potentials $V[z\,|\,{}^1_1\mathcal{G}^{101}_{\mathbf{t}_+,m}]$ are not ‘shape-invariant’ in the sense that the CLDT with the FF $\mathbf{c}$,0 does not preserve the order of the polynomial forming the appropriate AEH solution. It should be stressed that (as pointed in our works [11, 21 ] any shape-invariant GRef potential always retains its form under all the basic CLDTs and the CLDT with the normalizable FF is only one of these (generally four) transformations. Each basic CLDT converts any AEH solution into another AEH solution however the polynomials forming these solutions have different orders in general. Recently Quesne [55, 56] pointed to some examples of such polynomial mappings referring to them as ‘ shape-invariance of a new type’. As a matter of fact, these examples simply confirm that the preservation of the polynomial order under a basic CLDT is a nontrivial exclusion rather than the general rule.

## 7. Conclusions and further developments

It is shown that the regular-at-infinity solution of the 1D Schrödinger equation with the hyperbolic Pöschl-Teller (h-PT) potential (s–1)*sh*$^{-2}$r–(n–s+2)(n–s+1)*ch*$^{-2}$r, where s and n are positive integers, is expressible in terms of a n-order Heun polynomial $\mathrm{Hp}_n[y;\kappa;s]$ in

y≡ *th*r at an arbitrary negative energy $-\kappa^2$. It was proven that the Heun polynomials in question form a subset of generally complex Lambe-Ward polynomials corresponding to zero value of the accessory parameter. Since the mentioned solution expressed in the new variable y has an almost-everywhere holomorphic (AEH) form it can be used as the factorization function (FF) for ‘canonical Liouville-Darboux transformations’ (CLDTs) to construct a continuous family of ‘shape-invariant’ rational potentials $^1V[y;s,n|\kappa_1]$ exactly-solvable by the so-called ‘Hp-seed’ (HpS) Heine polynomials. There are also two ($\mathsf{t}_+ = \mathsf{a}$ or $\mathsf{a}'$) sequences of infinitely many rational potentials $^1V[y;s,n|\mathsf{t}_+,m]$ generated using CLDTs with nodeless regular-at-origin AEH FFs.

We point to three classes of SUSY partners of r-GRef potentials [20, 21, 11] which can be either exactly or conditionally-exactly[x)] quantized by Heun polynomials constructed in such a way. Each of these three classes will be studied in a separate publication.

## Appendix

## Basic ladder relations for regular solutions of the Schrödinger equation with theh-PT potential

In the particular case of the h-PT potential (a=0, b=1) parameters (3.40) and (3.40*) of the hypergeometric functions in the right-hand sides of(3.39a) and (3.39b) take the form

$$\alpha_{\pm,\sigma,\sigma}(-\kappa^2;\lambda_o,\mu_o;0)+\beta_{\pm,\sigma,\sigma}(-\kappa^2;\lambda_o,\mu_o;0)=1\pm\lambda_o+\sigma\kappa \qquad \text{(A.1)}$$

and

$$\beta_{\pm,\sigma,\sigma}(-\kappa^2;\lambda_o,\mu_o;0)-\alpha_{\pm,\sigma,\sigma}(-\kappa^2;\lambda_o,\mu_o;0)=\mu_o. \qquad \text{(A.1*)}$$

Let us first examine an effect of four basic DTs on the R@∞ solution

$$\Phi_{-,+,+}[y^2\,|\,_1\mathcal{G}^{101}]=\sqrt{2}\,y^{1\pm\lambda_o}\left(1-y^2\right)^{\frac{1}{2}(1+\kappa)} \times F[\tfrac{1}{2}(\kappa-t)-s+1,\tfrac{1}{2}(\kappa+t+1),\kappa+1;1-y^2], \qquad \text{(A.2)}$$

where we set

$$\alpha_{-,+,+}(-\kappa^2; s-{}^1\!/_2, t+s-{}^1\!/_2; 0) = {}^1\!/_2(\kappa - t - 2s + 2) = \kappa - \kappa^{s,t}_{\mathbf{b},0} \qquad \text{(A.3b)}$$

$$= s - {}^1\!/_2(\kappa + 1 - \kappa^{s,t}_{\mathbf{c},0}) \qquad \text{(A.3′)}$$

and

$$\beta_{-,+,+}(-\kappa^2; s-{}^1\!/_2, t+s-{}^1\!/_2; 0) = {}^1\!/_2(\kappa + t + 1) = {}^1\!/_2(\kappa + \kappa^{s,t}_{\mathbf{d},0}) \qquad \text{(A.3d)}$$

$$= s - {}^1\!/_2(\kappa^{s,t}_{\mathbf{a},0} + \kappa + 1) \qquad \text{(A.3″)}$$

taking advantage of (5.2a)-(5.2d). Coupled with two companion relations

$$\kappa + 1 - \alpha_{-,+,+}(-\kappa^2; s-{}^1\!/_2, t+s-{}^1\!/_2; 0) = {}^1\!/_2(\kappa + t + 2s) = {}^1\!/_2(\kappa + \kappa^{s,t}_{\mathbf{a},0}), \qquad \text{(A.3a)}$$

and

$$\kappa + 1 - \beta_{-,+,+}(-\kappa^2; s-{}^1\!/_2, t+s-{}^1\!/_2; 0) = {}^1\!/_2(\kappa - t + 1) = {}^1\!/_2(\kappa - \kappa^{s,t}_{\mathbf{c},0}), \qquad \text{(A.3c)}$$

we use the listed formulas for parameters of hypergeometric functions in the right-hand side of (A.2)to derive a set of the aforementioned basic ladder relations. In particular, an analysis of (A.3b) and (A.3d) immediately shows that either α- or β-coefficient of the hypergeometric function is unaffected by the CLDTs with the FFs $\mathbf{c}$,0 and $\mathbf{d}$,0 or $\mathbf{a}$,0 and$\mathbf{b}$,0, respectively, whereas the second coefficient changes by one. On other hand, the third parameter is unaffected by all four CLDTs. For this reason we start from the differential relations

$$\left\{\alpha - (1-z)\frac{d}{dz}\right\} F[\alpha, \beta, \gamma; 1-z] = \alpha F[\alpha+1, \beta, \gamma; 1-z] \qquad \text{(A.4)}$$

and

$$\left\{z(1-z)\frac{d}{dz} - \alpha z + \alpha + \beta - \gamma\right\} F[\alpha, \beta, \gamma; 1-z] = (\beta - \gamma) F[\alpha, \beta - 1, \gamma; 1-z] \qquad \text{(A.4*)}$$

which are obtained from the first two of the recurrence formulas (4.9.7) in [57] by changing z for 1-z. One can easily verify that any basic CLDT is represented in the space of the hypergeometric functions of our interest by either the generic ladder operators

$$ {}_1\hat{g}_{G;0}(\lambda,\nu) = z(1-z)\frac{d}{dz} + \tfrac{1}{2}[\lambda-(\lambda+\nu)z] \qquad \text{(A.5)} $$

for R@O FFs (type **a** or **c**) or by the so-called [11, 21] '*abridged*' ladder operators

$$ {}_1\hat{a}_{G;0}(\nu) = (1-z)\frac{d}{dz} - \tfrac{1}{2}\nu \qquad \text{(A.5*)} $$

for R@∞ FFs (type **b** or **d**). After setting $\Delta\lambda$ and $\Delta\nu$ to $-2\lambda_o$ and to $\kappa-\lambda_{1;\dagger,0}$, respectively, with the exponents $\lambda_{1;\dagger,0}$ defined via (5.2a)-(5.2d), we then substitute

resultant operators (A.5*) and (A.5) into the left-hand sides of (A.4*) and (A.4). This gives

$$ {}_1\hat{a}_{G;0}(\lambda,\nu)(\kappa-\kappa^{s,t}_{\mathbf{a},0})\,F[\tfrac{1}{2}(\kappa-t)-s+1,\tfrac{1}{2}(\kappa+t+1),1+\kappa;1-z] $$
$$ = -\tfrac{1}{2}(\kappa+t+2s)\,F[\tfrac{1}{2}(\kappa-t)-s,\tfrac{1}{2}(\kappa+t+1),\kappa+1;1-z], \qquad \text{(A.6a)} $$

$$ {}_1\hat{a}_{G;0}(\lambda,\nu)(\kappa-\kappa^{s,t}_{\mathbf{b},0})\,F[\tfrac{1}{2}(\kappa-t)-s+1,\tfrac{1}{2}(\kappa+t+1),1+\kappa;1-z] \qquad \text{(A.6b)} $$
$$ = -[\tfrac{1}{2}(\kappa-t)-s+1]\;F[\tfrac{1}{2}(\kappa-t)-s+2,\tfrac{1}{2}(\kappa+t+1);\kappa+1;1-z], $$

$$ {}_1\hat{g}_{G;0}(\lambda,\nu)(1-2s,\kappa-\kappa^{s,t}_{\mathbf{c},0})\,F[\tfrac{1}{2}(\kappa-t)-s+1,\tfrac{1}{2}(\kappa+t+1),\kappa+1;1-z] $$
$$ = -\tfrac{1}{2}(\kappa-t+1)\,F[\tfrac{1}{2}(\kappa-t)-s+1,\tfrac{1}{2}(\kappa+t-1),\kappa+1;1-z], \qquad \text{(A.6c)} $$

$$ {}_1\hat{a}_{G;0}(\kappa+\kappa^{s,t}_{\mathbf{d},0})F[\tfrac{1}{2}(\kappa-t)-s+1,\tfrac{1}{2}(\kappa+t+1),1+\kappa;1-z] $$

$$ = -\tfrac{1}{2}(\kappa+t+1)\,F[\tfrac{1}{2}(\kappa-t)-s+1,\tfrac{1}{2}(\kappa+t+3);\kappa+1;1-z]. \qquad \text{(A.6d)} $$

By expressing both operators (A.4) and (A.4*) in terms of y:

$$ {}_1\hat{g}_{G;0}(\lambda,\nu) = \tfrac{1}{2}y(1-y^2)\frac{d}{dy} + \tfrac{1}{2}[\lambda-(\lambda+\nu)y^2] \qquad \text{(A.7)} $$

and

$$y_1 \hat{a}_{G;0}(\nu) = \tfrac{1}{2}\left[(1-y^2)\frac{d}{dy} - \nu y\right] \tag{A.7*}$$

and taking into account that

$$\alpha_{-,+,+}(-\kappa^2; s-\tfrac{1}{2}, t+s-\tfrac{1}{2}; 0) \pm 1 = \alpha_{-,+,+}(-\kappa^2; s\mp 1-\tfrac{1}{2}, t+s\mp 1-\tfrac{1}{2}; 0) \tag{A.8}$$

and

$$\beta_{-,+,+}(-\kappa^2; s-\tfrac{1}{2}, t+s-\tfrac{1}{2}; 0) \pm 1 = \beta_{-,+,+}(-\kappa^2; s-\tfrac{1}{2}, t\pm 2+s-\tfrac{1}{2}; 0) \tag{A.8*}$$

one can then convert (A.6a)-(A.6d) into the recurrence formulas for Heun functions (4.31):

$$\left[y(1-y^2)\frac{d}{dy} - 2s+1-(\kappa+t+1)y^2\right]\mathrm{Hf}[y;\kappa;t,s;-,+,+] = -(\kappa+t+2s)\mathrm{Hf}[y;\kappa;t,s+1;-,+,+], \tag{A.9a}$$

$$\left[(1-y^2)\frac{d}{dy} - (\kappa-t-2s+2)y\right]\mathrm{Hf}[y;\kappa;t,s;-,+,+] = (t+2s-2-\kappa)\,y\,\mathrm{Hf}[y;\kappa;t-2,s+1;-,+,+], \tag{A.9b}$$

$$\left[y(1-y^2)\frac{d}{dy} - 2s+1-(\kappa-t-2s+2)y^2\right]\mathrm{Hf}[y;\kappa;t,s;-,+,+] = (t-1-\kappa)\mathrm{Hf}[y;\kappa;t,s+1;-,+,+], \tag{A.9c}$$

$$\left[(1-y^2)\frac{d}{dy} - (\kappa+t+1)\,y\right]\mathrm{Hf}[y;\kappa;t,s;-,+,+] = -(\kappa+t+1)y\,\mathrm{Hf}[y;\kappa;t+2,s-1;-,+,+] \tag{A.9d}$$

In case of R@∞ solutions we can also write a similar set of basic recurrence formulas for the Heun functions (4.30):

$$\left[y(1-y^2)\frac{d}{dy} - (t+1)y^2 - \kappa y - 2s+1\right]\mathrm{Hf}[y;\kappa;t,s;-,+,-] = -(\kappa+t+s)\,\mathrm{Hf}[y;\kappa;t,s+1;-,+,-], \tag{A.10a}$$

$$\left[(1-y^2)\frac{d}{dy}+(t+2s-2)y-\kappa\right]\mathrm{Hf}[y;\kappa;t,s;-,+,-] \tag{A.10b}$$

$$=(t+2s-2-\kappa)\;y\,\mathrm{Hf}[y;\kappa;t,s-1;-,+,-],$$

$$\left[y(1-y^2)\frac{d}{d\varsigma}+(t+2s-2)y^2-\kappa y-2s+1\right]\mathrm{Hf}[y;\kappa;t,s;-,+,-] \tag{A.10c}$$

$$=(t-1-\kappa)\,\mathrm{Hf}[y;\kappa;t-2,s+1;-,+,-],$$

$$\left[(1-y^2)\frac{d}{dy}-(t+1)y-\kappa\right]\mathrm{Hf}[y;\kappa;t,s;-,+,-] \tag{A.10d}$$

$$=-(\kappa+t+1)\;y\,\mathrm{Hf}[y;\kappa;t+2,s-1;-,+,-].$$